\documentclass{article}

\usepackage{amssymb,amsfonts,amsmath}
\usepackage{cite,enumerate,float,indentfirst}
\usepackage{color}

\def\be{\begin{eqnarray}}
\def\ee{\end{eqnarray}}
\def\nn{\nonumber}

\def\p{\partial}

\def\Tr{{\rm Tr}\,}

\def\RCM{RCM\ }

\def\BBigl{{\Big<\!\!\!\Big<}}
\def\BBigr{\Big>\!\!\!\Big>}
\def\tchi{\tilde\chi}

\definecolor{red}{rgb}{1,0,0}
\definecolor{orange}{rgb}{1,0.5,0}
\definecolor{violet}{rgb}{0.7,0,1}

\def\cre{\color{red}}

\def\cg{\color{green}}
\def\cb{\color{blue}}

\hoffset= - 1.0in         % switch off for draft style
\begin{document}

\title{\vspace{1.1cm}{\LARGE {\bf Tensorial generalization of characters
}\vspace{.5cm}}
\author{{\bf H. Itoyama$^{a}$},
{\bf A. Mironov$^{b,c,d}$},
\ {\bf A. Morozov$^{e,c,d}$}
}
\date{ }
}

\maketitle

\vspace{-6.2cm}

\begin{center}
\hfill FIAN/TD-17/19\\
\hfill ITEP/TH-27/19\\
\hfill IITP/TH-18/19\\
\hfill MIPT/TH-16/19\\
\hfill OCU-PHYS-506\\
\hfill NITEP 29
\end{center}

\vspace{3.cm}

%\begin{center}
%$\left.\text{\parbox{11cm}{$^a$ {\small {\it Nambu Yoichiro Institute of Theoretical and Experimental Physics (NITEP)}}\\
%$^b$ {\small {\it \it Department of Mathematics and Physics, Graduate School of Science}}\\
%$^c$ {\small {\it Osaka City University Advanced Mathematical Institute (OCAMI)}}}}\right.$:
%\parbox{5.4cm}{\small {\it Osaka City University, 3-3-138, Sugimoto, Sumiyoshi-ku, Osaka, 558-8585, Japan}}\\
%\end{center}
\vspace{-.27cm}

\noindent
$^a$ {\small {\it Nambu Yoichiro Institute of Theoretical and Experimental Physics (NITEP) and}} \\
\phantom{a} {\small {\it Department of Mathematics and Physics, Graduate School of Science,}}  \\
\phantom{a} {\small {\it Osaka City University, 3-3-138, Sugimoto, Sumiyoshi-ku, Osaka, 558-8585, Japan  }}\\
$^b$ {\small {\it I.E.Tamm Theory Department, Lebedev Physics Institute, Leninsky prospect, 53, Moscow 119991, Russia}}\\
$^c$ {\small {\it ITEP, B. Cheremushkinskaya, 25, Moscow, 117259, Russia }}\hfill\\
$^d$ {\small {\it Institute for Information Transmission Problems,  Bolshoy Karetny per. 19, build.1, Moscow 127051 Russia}}\\
$^e$ {\small {\it MIPT, Dolgoprudny, 141701, Russia}}\\

\vspace{.5cm}

\begin{abstract}
In rainbow tensor models,
which generalize rectangular complex matrix model (RCM) and possess
a huge gauge symmetry $U(N_1)\times\ldots\times U(N_r)$,
we introduce a new sub-basis in the linear space of gauge invariant operators, which is a redundant basis in the space of operators with non-zero Gaussian averages.
Its elements are labeled by $r$-tuples of Young diagrams of a given size
equal to the power of tensor field.
Their tensor model averages are just products
of dimensions:\ $\Big<\chi_{R_1,\ldots,R_r}\Big> \sim C_{R_1,\ldots, R_r}
D_{R_1}(N_1)\ldots D_{R_r}(N_r)$ of representations $R_i$ of the linear group $SL(N_i)$,
with $C_{R_1,\ldots, R_r}$ made of the Clebsch-Gordan coefficients of representations $R_i$ of the symmetric group.
Moreover, not only the averages, but the operators $\chi_{\vec R}$ themselves
exist only when these $C_{\vec R}$ are non-vanishing.
This sub-basis is much similar to the basis of characters (Schur functions)
in matrix models, which is
distinguished  by the property $\Big<{\rm character}\Big> \sim { character}$,
which
opens a way to lift the notion and the theory of characters (Schur functions)
from matrices to tensors.
In particular,
operators $\chi_{\vec R}$ are eigenfunctions of operators
which generalize the usual cut-and-join operators $\hat W$; they satisfy orthogonality conditions similar to the standard characters, but they do not form a {\it full} linear basis for all gauge-invariant
operators, only for those which have non-vanishing Gaussian averages.
\end{abstract}

\bigskip

\bigskip

\section{Introduction}

Tensor models \cite{tenmods} look {\it super}integrable \cite{IMMarist}
and can be exactly solved \cite{rainbow,MMtenmod}
by combinatorial methods \cite{Ramg1,MMtenmod}
just like their matrix model prototypes \cite{MMcharMAMO}.
However, original presentations
\cite{Ramg1,MMtenmod}
rely too much on the theory of symmetric groups
and therefore can not attract much attention from physicists,
generically unfamiliar with this subject.
The goal of this note is to provide another formulation,
which is much simpler.
Following the strategy of \cite{rainbow,IMMarist},
we concentrate on the Gaussian averages,
i.e. apply the standard functional integral technique of
quantum field theory.
As often happens, this eliminates unnecessary details
and provides clear statements,
while details can be restored afterwards.
In this particular case, the main source of sophistication
is the complicated structure of the space of gauge-invariant
operators, but in fact many of them have vanishing Gaussian averages.
Instead, the space of those with non-vanishing averages turns to be
very simple, and it has a basis very similar to the
basis of ordinary Schur functions.
Thus, despite one does not expect straightforward applicability
of the ordinary group theory in the tensor case,
a very important sub-sector is actually controlled by something
of this kind and can be investigated in depth.
An option for the remaining part of the Hilbert space
is a kind of non-linear realization as a qualitatively new
{\it extension} of the idea of the separation into
single- and multi-trace operators (while
literally in the formalism of the present paper, the multi-trace operators
for $r=2$ are treated on equal footing with the single-trace ones).

\bigskip

Our main idea in this paper is to concentrate on lifting to the tensor level of the main feature of matrix models,
which,  according to \cite{rainbow,MMcharMAMO}, is that the average of linear group character in representation $R$ (the Schur function) is again a character:
\be
\Big< \chi_R \Big> \sim \chi_R^*
\label{charave}
\ee
where  the argument of character at the l.h.s.
is the matrix-integration variable,
while the character at the r.h.s. is taken on the
``topological locus".
Concrete expressions depend on the {\it model}
(i.e. on the choice of the Vandermonde weight in the measure),
and the suppressed $R$-dependent coefficient at the r.h.s.,
on its {\it phase}
(i.e. the choice of the exponentiated Casimir weights in the measure,
see \cite{DV,AMM} for the important notion of the {\it phase} in
the theory of matrix-model and functional integrals).
For examples of the different models in the Gaussian phase,
see \cite{MMcharMAMOreview,Shaqt}. For examples beyond Gaussian phase, see \cite{PSh}.

In this paper, we start with the rectangular complex matrix model (\RCM) \cite{MMMM},
which has a straightforward generalization to {\it rainbow} \cite{rainbow}
tensor models \cite{tenmods}-\cite{MMtenmod},
with well-studied {\it Aristotelian} (rank $r=3$) model \cite{IMMarist}
as the first non-trivial example.
Our goal is to {\it pose} the problem of how (\ref{charave}) is generalized
to $r>2$ and suggest a possible way to solve it.
Deeper questions such as generalization of the group theory,
which could underline the emerging structures
are yet too early to address, and we leave them beyond the scope of the present paper.
To avoid unnecessary complications, we consider only the Gaussian phase.

In the case of  \RCM, eq.(\ref{charave}) can be written in full detail as
\be
\!\!\!\!
\frac{{\int_{N_1\times N_2}}  \chi_R\{P_k=\Tr(M\bar M)^k\} e^{-\Tr M\bar M} d^2M}
{\int_{N_1\times N_2}    e^{-\Tr M\bar M} d^2M}
= \boxed{\Big< \chi_R\{P_k\}\Big>
=  \frac{D_R(N_1)D_R(N_2)}{d_R}
}
= \frac{\chi_R\{p_k=N_1\}\chi_R\{p_k=N_2\}}{d_R}
\label{RCMaverage}
\ee
where $\chi_R$ is the Schur function, which is a symmetric function of some variables $x_i$ (here the eigenvalues of the matrix $M\bar M$) or a function of time variables $p_k:=\sum_i x_i^k$, $d_R=\chi_R(p_k=\delta_{1,k})$.
The integrals here are over $N_1\times N_2$ (rectangular) complex matrices $M$,
the model has a ``gauge" symmetry $U(N_1)\times U(N_2)$.
The character (Schur function) depends on the Young diagram $R$
and on the sequence of time-variables $p_k$, e.g.
$$
\chi_\emptyset\{p\} = 1, \ \ \ \chi_{[1]}\{p\} = p_1, \ \ \
\chi_{[2]}\{p\} = \frac{p_2+p_1^2}{2}, \ \ \ \chi_{[1,1]}\{p\} = \frac{-p_2+p_1^2}{2}, \ \ \
\ldots
$$
If one ascribes $p_k$ the grading $k$, then $\chi_R\{p\}$ is homogeneous of degree $|R|$,
which is the size (number of boxes) of the diagram $R$.
The role of time-variables at the l.h.s. in eq.(\ref{RCMaverage}) is played by
the single-trace gauge invariants $P_k\equiv \Tr (M\bar M)^k$,
while at the r.h.s. they are fixed at the ``classical topological locus",
where all $p_k=N$. At these values, the character of the representation $R$ is equal to the dimension of the representation $R$, $D_R(N)$.

Now comes the {\bf first crucial observation}:
bilinear combination is {\it not} generic,
from the entire set of all $\chi_{R_1}\{N_1\}\chi_{R_2}\{N_2\}$,
only the subset $R_2=R_1$ is selected.
If we look at the result of \cite{IMMarist} for Aristotelian model,
we easily observe that the same is true there:
the set of averages of a given grading is not an arbitrary function of colorings
$N_1, N_2, N_3$, but is restricted.
Say, at level $n=2$, allowed at the r.h.s. are only
$$
\chi_{[2],[2],[2]}^* \ \ \ \chi_{[2],[1,1],[1,1]}^* \ \ \ \chi_{ [1,1],[2],[1,1]}^*
\ \ \ \chi_{[1,1],[1,1],[2]}^*
$$
while the other four
$$\chi_{[2],[2],[1,1]}^* \ \ \ \chi_{[2],[1,1],[2]}^* \ \ \ \chi_{ [1,1],[1,1],[2]}^*
\ \ \ \chi_{[1,1],[1,1],[1,1]}^*
$$
are forbidden. We introduced here an abbreviated
notation $\chi_{_{R_1,R_2,R_3}}^* = D_{R_1}(N_1)D_{R_2}(N_2)D_{R_3}(N_3)$.
The {\bf second  observation} is that the allowed $\chi_{R_1,R_2,R_3}^*$ are actually
averages of the operators $\chi_{R_1,R_2,R_3}$ which are of the symmetry type
$R_1\otimes R_2\otimes R_3$ from the point of view of the group $S_n^{\otimes 3}$.
Finally, the {\bf third observation} is that an attempt to build up an operator
$\chi_{R_1,R_2,R_3}$ of forbidden symmetry type gives zero,
this is the reason why such $\chi_{R_1,R_2,R_3}$ does not emerge among averages.

Our goal in this text is to study and extend these observations to other
gradings and ranks
with an obvious purpose to understand the way to generalize (\ref{charave})
and, hence, the very notion of character (Schur functions) to tensor models.
Though looking at the Gaussian averages is simpler than at the operators, we emphasize that the story is actually about
operators ${\cal K}$,
and, like in the matrix model case, it is essentially independent of the
averaging procedure, in particular, of the choice of the Gaussian phase of the model.

{\bf The  main result} of the present paper is that (\ref{RCMaverage})
has a direct generalization to the rainbow tensor models:
there is a set of gauge invariant operators
\be
\boxed{
\chi_{R_1,\ldots, R_r}(M,\bar M)
= \frac{1}{n!}\!\!\!\!\!\!\!\!\!\sum_{\ \ \ \ \ \sigma_1,\ldots,\sigma_r\in S_n}\!\!\!\!\!\!\!\!\!
\psi_{R_1}(\sigma_1)\ldots\psi_{R_r}(\sigma_r)\cdot {\cal K}^{(n)}_{\sigma_1,\ldots,\sigma_r}
}
\label{hurchar}
\ee
which are linear combinations of the tensorial counterparts of ``multi-trace" operators
\be
{\cal K}^{(n)}_{\sigma_1,\ldots,\sigma_r}
= \sum_{\vec a^1=1}^{N_1}\ldots\sum_{\vec a^r=1}^{N_r}
\left(\prod_{p=1}^n M_{a^1_p,...a^r_p}
\bar M^{a^1_{\sigma_1(p)},\ldots,a^r_{\sigma_r(p)}}\right)
\ee
with the coefficients made from symmetric group characters $\psi_R(\sigma)$
such that their Gaussian averages with the weight
$\exp\Big(-{\cal K}^{(1)}_{id,\ldots,id}\Big)
= \exp\left(-\sum_{a^1=1}^{N_1}\ldots \sum_{a^r=1}^{N_r}
M_{a^1\ldots a^r}\bar M^{a^1\ldots a^r}\right)$
are essentially the products of dimensions:
\be\label{Ga}
\boxed{
\Big< \chi_{R_1,\ldots, R_r} \Big> \ =
\ C_{R_1,\ldots,R_r} \cdot
\frac{D_{R_1}(N_1)\cdot \ldots\cdot D_{R_r}(N_r)}{d_{R_1}\cdot\ldots \cdot d_{R_r}}
}
\ee
The sizes of all the Young diagrams $R_k$ are the same and equal to the power $n$ in $M$ and $\bar M$,
which we call {\it level} in what follows.
Our main claim is that $\chi_{\vec R}$ form the full  basis of operators
with non-vanishing Gaussian averages (see sec.\ref{nvG}).
This basis is overcomplete, i.e. there are linear dependencies, but it is smaller
than the full basis of {\it all} gauge invariant operators, which grows much faster
than the number of $\chi_{\vec R}$ and even than the $r$-th power
of the number of Young diagrams. The reason is that there are many operators with
vanishing Gaussian averages.
The coefficients $C_{\vec R}$ are $r$-counterparts of the Clebsch-Gordan coefficients,
which, in fact, just select allowed symmetries.
In the case of $r=2$, i.e. for the RCM, $C_{R_1,R_2} = d_{R_1}\delta_{R_1,R_2}$
and we return to (\ref{RCMaverage}).
Actually, $\chi_{\vec R}$ itself is proportional to $C_{\vec R}$,
i.e. it is the operator, not just the average, which vanishes in the case
of forbidden symmetry.

Perhaps, a main problem with $\chi_{\vec R}$ is that, for $r>2$, they are too few,
much less than the number of gauge-invariant operators, which are labeled by a peculiar
double coset of symmetric group $S_n\backslash S_n^r/S_n$ \cite{IMMarist}.
Thus one can continue a search for more genuine {\it tensorial characters},
or at least for the {\it coset} ones.
Unlike $\chi_{\vec R}$, which deserve the name {\it Kronecker characters},
generic tensorial and coset ones are not expected to be invariant under arbitrary conjugations:
the symmetry of the coset is smaller.
We do not go into details of this further generalization,
the story of Kronecker characters $\chi_{\vec R}$ is already
quite something to consume: an unexpected and far going generalization from
matrices to tensors, see also \cite{IMMbr}.

Our main target in this paper is the Kronecker tensorial characters (\ref{hurchar})
and their properties.
We start in sec.2 with explaining the main idea in simple examples of first levels of the RCM and the Aristotelian $r=3$ model. Then, in sec.3, we introduce formal definition of the generalized (Kronecker) characters and discuss their properties. Sec.4 is devoted to the orthogonality of these characters, while, in sec.5, we introduce the generalized cut-and-join operators $\hat W$ that have the generalized characters as their eigenvalues. Sec.6 contains some concluding remarks.

\section{The main idea}

\subsection{Operators with a given symmetry}

Our main idea is to consider the basis of gauge invariant operators associated with elements of the tensor product of $r$ copies of the group algebras of symmetric group.

Consider an element of the group algebra of $S_n$,
\be
\hat R_\alpha = \sum_{\sigma\in S_n} \alpha_\sigma \cdot \hat\sigma
\ee
where $\sigma$ are the elements of the symmetric group $S_n$.
We now construct a basis in the space of operators ${\cal K}$ of level $n$,
by the action of $r$ operations $\hat R_1\otimes \ldots \otimes \hat R_r$
on the operator ${\cal K}_{_{id^r}}^n$.
For instance, for $r=3$
\be
\hat R_\alpha\otimes \hat R_{\alpha'} \otimes \hat R_{\alpha''} =
\sum_\sigma \alpha_\sigma \hat\sigma \otimes
\sum_\sigma \alpha'_\sigma  \hat\sigma \otimes
\sum_\sigma \alpha''_\sigma  \hat\sigma
\ee
acts on the operator ${\cal K}_{_{id,id,id}}^n=\prod_{p=1}^n M_{a_pb_pc_p}\bar M^{a_pb_pc_p} $ as follows:
\be
\hat R_\alpha\otimes \hat R_{\alpha'} \otimes \hat R_{\alpha''}:\ \ \ \ {\cal K}_{_{id,id,id}}^n \longrightarrow\chi_{\alpha,\alpha',\alpha''}=
\sum_{\sigma,\sigma',\sigma''\in S_n} \alpha_\sigma\alpha'_{\sigma'}\alpha''_{\sigma''}
\prod_{p=1}^n M_{a_pb_pc_p}\bar M^{a_{\sigma(p)}b_{\sigma'(p)}c_{\sigma''(p)}}
\ee
Now one can choose the coefficients $\alpha$'s in such a way that they are associated with some symmetry patterns described by Young diagrams: with representations of $S_n$. For instance, up to the level $n=3$, there are 3 different patterns:
\begin{itemize}
\item{$[n]\in S_n$:} symmetrization with all the equal weights of one unit
\be
\hat S = \sum_{\sigma\in S_n} \hat\sigma
\ee
\item{$[1^n]\in S_n$:} antisymmetrization with  weights depending on the parity of
the permutation
\be
\hat A = \sum_{\sigma\in S_n} (-)^{P_\sigma}\hat\sigma
\ee
\item{$\left[2,1\right]\in S_3$:} in this particular case we {\it define}
\be
\hat B = 1-\hat P_{13}  +\hat P_{12} - \hat P_{13}\hat P_{12}
\ee
This definition follows the standard rule:
make a Young tableau
$\begin{array}{|c|c|} \hline   1 & 2 \\ \hline     3 \\  \cline{1-1} \end{array}$,
first symmetrize in rows and then antisymmetrize in columns:
\be
\hat B \psi^{a_1a_2a_3}  = \psi^{a_1a_2a_3}+ \psi^{a_2a_1a_3}
- \psi^{a_3a_2a_1} - \psi^{a_2a_3a_1}
\ee
One could start from another Young tableau
$\begin{array}{|c|c|} \hline   1 & 3 \\ \hline     2  \\  \cline{1-1} \end{array}$
and obtain instead
\be
\hat {\tilde B} = 1-\hat P_{12}  +\hat P_{13} - \hat P_{12}\hat P_{13}
\nn
\ee
\be
\hat {\tilde B}  \psi^{a_1a_2a_3} = \psi^{a_1a_2a_3}+ \psi^{a_3a_2a_1}
- \psi^{a_2a_1a_3} -\psi^{a_3a_1a_2}
\ee
\end{itemize}

\subsection{RCM}

Let us start with the $r=2$ case. If we consider averages (\ref{RCMaverage}) of a given grading, we obtain functions of $N_1$ and $N_2$,
which are inhomogeneous, but not arbitrary:

\begin{itemize}

\item{Level 1:}  ${\rm Im}_1^{(2)} = {\rm Span}\{D_{[1]}(N_1)D_{[1]}(N_2) = N_1N_2\}$

\item{Level 2:}  ${\rm Im}_2^{(2)} = {\rm Span}\{D_{[2]}(N_1)D_{[2]}(N_2)=N_1N_2(N_1+1)(N_2+1),\
\ D_{[1,1]}(N_1)D_{[1,1]}(N_2) =N_1N_2(N_1-1)(N_2-1)\}$.
However, neither $D_{[2]}(N_1)D_{[1,1]}(N_2) = N_1N_2(N_1+1)(N_2-1)$ nor $D_{[1,1]}(N_1)D_{[2]}(N_2) =
N_1N_2(N_1-1)(N_2+1)$ belongs to ${\rm Im}_2^{(2)}$.

\item{Level $n$:} The same persists at higher levels:
 ${\rm Im}_n^{(2)} = {\rm Span}\{D_{R}(N_1)D_{R}(N_2), \ R\vdash n$\},
but none of non-diagonal $D_{R_1}(N_1)D_R(N_2)$ with $R_1,R_2\vdash n$ and
$R_2\neq R_1$ belongs to {\it this} space.

\end{itemize}

\noindent
We used here the standard notation $R\vdash n$, meaning that the size
(number of boxes) in $R$ is $n$, i.e. $|R|=n$.
Dimension of ${\rm Im}_n^{(2)}$ is just the number of Young diagrams of the size $n$,
i.e.
\be
\sum_n {\rm dim}({\rm Im}_{n}^{(2)}) \cdot q^n =
\prod_{n} \frac{1}{1-q^n}
\label{dimIm2}
\ee

\bigskip

The fact that the image is spanned by a given set labeled with $R_1=R_2=R$,
means that one can enumerate and classify the {\it averaged} quantities:
they are labeled by Young diagrams, and, indeed, they are just characters.
Looking at the {\it averages}, one can {\it determine} that any
product of characters is linearly expressed through characters themselves.
This is a well known fundamental fact, but, what is important,
one can now extract it from studying the  averages,
and this approach can be straightforwardly extendable from matrices
to tensors, where we know neither what the characters are,
nor their properties.
Information that we need is just the structure of spaces
${\rm Im}_{n}^{(r)}$.
Moreover, one can begin just from the linear space
structure, namely from the basis in ${\rm Im}_{n}^{(r)}$
made from $r$-linear combinations of dimensions
\be
\chi^*_{\vec R}=\prod_{i=1}^r D_{R_i}(N_i)
\ee
with all $R_i\vdash n$.

In these terms, the statement for RCM is that
${\rm Im}_n^{(2)}$ is spanned by the ``diagonal" $\chi^*_{R,R}$,
and these are averages of $\chi_R\{P\}$ with $P_k=\Tr(M\bar M)^k$.

More than that, if we ask what could be the quantity,
whose average would produce a non-diagonal $\chi_{R_1,R_2}^*$,
the answer will be zero, this is why such averages
do not actually arise.
Indeed,  operators in the RCM can be {\it a priori} labeled by
two permutations $\sigma_1,\sigma_2\in S_n$:
\be\label{rcm}
{\cal O}_{\sigma_1,\sigma_2} =
\prod_{p=1}^n M_{a_pb_p}\bar M^{a_{\sigma_1(p)}b_{\sigma_2(p)}}
\ee
(summation is assumed over all $a_p$ and $b_p$).
However, not all these operators are independent: in fact
\be
n=1 &
{\cal O}_{(1),(1)} = \Tr M \bar M = P_1 \nn\\
n=2 &
{\cal O}_{(1)(2),(1)(2)} = {\cal O}_{(12),(12)} = P_1^2, \ \ \ \
{\cal O}_{(1)(2),(12)} =  {\cal O}_{(12),(1)(2)} = P_2  \nn\\
\ldots
\ee
so that for $S=(1)(2)+(12)=I+P $ and $A=(1)(2)-(12)=I-P$
\be
{\cal O}_{SS}=2(P_2+P_1^2) = 4\chi_{[2]}\{P\}, \ \ \ \
{\cal O}_{AA}=2(-P_2+P_1^2) = 4\chi_{[11]}\{P\}
\ee
while
\be
{\cal O}_{SA} = {\cal O}_{AS} = 0
\ee
In other words, the fact about the averages can be observed
{\it before} averaging, i.e. is actually independent of the {\it phase}
of the model.
Thus we think that technically the simplest approach to study the space of gauge invariant operators is provided by the expression of Gaussian averages through
dimensions $D_R(N)$.

\subsection{Aristotelian model
\label{levlevArist}}

In fact, all the information needed in this case can be extracted
from the detailed study made in \cite{IMMarist}.
We just need to reformulate these results in terms
which are relevant to purposes of the present paper.
We refer to \cite{IMMarist} for the tables of operators and
the notation.

\begin{itemize}

\item{Level 1:} \ ${\rm dim}({\rm Im}_{1}^{(3)})=1$, and the relevant character is $\chi_{[1],[1],[1]}^*=
D_{[1]}(N_1)D_{[1]}(N_2)D_{[N_3]}=N_1N_2N_3$,
the corresponding operator
${\cal K}_1 =\sum_{a=1}^{N_1}\sum_{b=1}^{N_2}\sum_{c=1}^{N_3} M_{abc}\bar M^{abc}
= M_{abc}\bar M^{abc}$
(hereafter, the summation over repeated indices $a,b,c,\ldots$
is assumed).

\item{Level 2:} \ ${\rm dim}({\rm Im}_{2}^{(3)})=4$, and the relevant characters are
\be
\chi_{SSS}^*=D_{[2]}(N_1)D_{[2]}(N_2)D_{[2]}(N_3)
= \frac{1}{8}N_1N_2N_3(N_1+1)(N_2+1)(N_3+1)
& = \frac{1}{8}\Big< {\cal K}_1^2+{\cal K}_{\cre 2}+ {\cal K}_{\cg 2} + {\cal K}_{\cb 2}\Big>\nn \\
\chi_{SAA}^*=D_{[2]}(N_1)D_{[1,1]}(N_2)D_{[1,1]}(N_3)
= \frac{1}{8}N_1N_2N_3(N_1+1)(N_2-1)(N_3-1)
& = \frac{1}{8}\Big< {\cal K}_1^2+{\cal K}_{\cre 2}-{\cal K}_{\cg 2} - {\cal K}_{\cb 2}\Big>\nn \\
\chi_{ASA}^*=D_{[1,1]}(N_1)D_{[2]}(N_2)D_{[1,1]}(N_3)
= \frac{1}{8}N_1N_2N_3(N_1-1)(N_2+1)(N_3-1)
& = \frac{1}{8}\Big< {\cal K}_1^2-{\cal K}_{\cre 2}+{\cal K}_{\cg 2}- {\cal K}_{\cb 2}\Big>\nn \\
\chi_{AAS}^*=D_{[1,1]}(N_1)D_{[1,1}(N_2)D_{[2]}(N_3)
= \frac{1}{8}N_1N_2N_3(N_1-1)(N_2-1)(N_3+1)
& = \frac{1}{8}\Big< {\cal K}_1^2-{\cal K}_{\cre 2}- {\cal K}_{\cg 2} +{\cal K}_{\cb 2} \Big>
\nn
\ee
with $S=[2]$ and $A=[1,1]$. The operators ${\cal K}$ are defined in \cite[eq.(7.8) and Appendix A.2]{IMMarist}.
Thus, ${\rm Im}_{2}^{(3)}$ is spanned by $\chi_{SSS}^*$ and $3\times\chi_{SAA}^*$,
where $3$ stands for the number
 of quantities obtained by permutations from $S_3$.
Forbidden are the four other characters
\be
\chi_{SSA}^*=D_{[2]}(N_1)D_{[2]}(N_2)D_{[1,1]}(N_3)=
\frac{1}{8}N_1N_2N_3(N_1+1)(N_2+1)(N_3-1),
\ee
$ \chi_{SAS}^*, \ \chi_{ASS}^*$ and $\chi_{AAA}^*$:
there are no such operators at level $2$ of the Aristotelian model
with appropriate discrete  symmetry group properties.

\item{Level 3:} \  ${\rm dim}({\rm Im}_{3}^{(3)})=11$, and the relevant operators are
\be
\chi_{SSS}, \ 3\times \chi_{SAA}, \ 3\times \chi_{SBB}, \ 3\times \chi_{ABB}, \  \chi_{BBB}
\ee
 with $S=[3]$, $B=[2,1]$, $A=[1,1,1]$.
Forbidden   are
$ 3\times \chi_{SSA}$,  $3\times\chi_{SSB}$, $3\times\chi_{AAB}$, $6\times\chi_{SBA}$
and $\chi_{AAA}$.

Again, from analysis of averages, one obtains that, for instance,
\be
\chi_{SSS}^* \sim \Big<{\cal K}_1^3
+3({\cal K}_{\cre 2}+{\cal K}_{\cg 2}+{\cal K}_{\cb 2}){\cal K}_1
+2({\cal K}_{\cre 3}+{\cal K}_{\cg 3}+{\cal K}_{\cb 3})
+ 6({\cal K}_{{\cre 2},{\cg 2}}+ {\cal K}_{{\cre 2},{\cb 2}} + {\cal K}_{{\cg 2},{\cb 2}})
+ 2{\cal K}_{3W}\Big>
\ee
and the operator at the r.h.s. is exactly the triple symmetrization
of $M_{a_1b_1c_1}M_{a_2b_2c_2}M_{a_3b_3c_3}\bar M^{a_1b_1c_1}\bar M^{a_2b_2c_2}\bar M^{a_3b_3c_3}$
w.r.t. indices $(a_1,a_2,a_3)$,
$(b_1,b_2,b_3)$ and $(c_1,c_2,c_3)$ of $\bar M$ (they are defined in \cite[eq.(7.19) and Appendix A.3]{IMMarist}).
Forbidden symmetrizations such as $SBA$ are vanishing already at the operator level.

\end{itemize}

The $r=3$ counterpart of (\ref{dimIm2}),
\be
\sum_n {\rm dim}({\rm Im}_{n}^{(3)}) \cdot q^n =
\prod_{n=1} \frac{1}{(1-q^n)^{\#_{\rm dde(n)}}} = \nn \\ =
\prod_{n=1} \frac{1}{(1-q)(1-q^2)^{^3}(1-q^3)^{^7}(1-q^4)^{^{26}}(1-q^5)^{^{97}}
(1-q^6)^{^{624}}(1-q^7)^{^{4163}}\ldots}
\label{dimIm3}
\ee
%2, 15 (5???), 20, 107
appears to imply that the classification problem is
hopeless.
In fact, it was already addressed in \cite{Ramg1} and \cite{IMMarist},
but in what follows we look at it from a somewhat different direction in the spirit of \cite{Ramg2} and \cite{IMMbr}.

Note that the coefficients in (\ref{dimIm3}) grow much faster than the triples
of Young diagram (the cubes of the coefficients in (\ref{dimIm2})). This means that $\chi_{R_1,R_2,R_3}$ do not exhaust all gauge invariant operators. At the same time, as discovered in \cite{IMMarist},
the number of linearly independent {\it Gaussian} averages is lower
than the prediction of (\ref{dimIm3}), and it is actually less than the
number of triples. This is why $\chi_{R_1,R_2,R_3}$ are enough to enumerate all operators with non-vanishing Gaussian average, though $\chi_{R_1,R_2,R_3}$ form an overfull basis. This is reflected in the fact that the triple products of dimensions are not all linearly independent. The differences between the number of all gauge invariant operators, of $\chi_{R_1,R_2,R_3}$ and of independent Gaussian averages all appear starting from level $n=4$.

\begin{itemize}
\item{Level $n=4$:} Not all of the $43$ linearly independent operators at this level
have independent Gaussian averages:
this was overlooked in \cite{IMMarist}, where the phenomenon was first
observed only at level 5.
In fact, independent are just $30$ averages, while there are relations
\be
<{\cal K}_{3W}{\cal K}_1> +3<{\cal K}_{222}> + 2<{\cal K}_{\cg 2}^2>
\ = 2<{\cal K}_{\cg 3}{\cal K}_1>
+ 2<{\cal K}_{{\cg 2},{\cre 2},{\cb 2}}>+ 2<{\cal K}_{{\cg 2},{\cb 2},{\cre 2}}>
\nn \\
<{\cal K}_{\cre 31W}> \ = \ <{\cal K}_{{\cg 2},{\cb 2}}{\cal K}_1>
-<{\cal K}_{\cg 2}{\cal K}_{\cb 2}> + <{\cal K}_{\cre 22W}>
\label{lev4rels3}
\ee
plus four more relations obtained by cyclic permutations of colorings (these operators are defined in \cite[eq.(7.26) and Appendix A.4]{IMMarist}).
Thus, for instance, the combination
\be
{\cal K}_{\cre 31W} - {\cal K}_{{\cg 2},{\cb 2}}{\cal K}_1
+{\cal K}_{\cg 2}{\cal K}_{\cb 2} - {\cal K}_{\cre 22W}
\label{licochi4}
\ee
is a non-trivial operator, but its average vanishes.
The $30$-dimensional linear space of averages is spanned by $43$
triple-characters $\chi_{R_1,R_2,R_3}^*$. They also produce an overfull basis in the space of operators with non-vanishing Gaussian averages.
\end{itemize}

\bigskip

We make a table for the Aristotelian model ($r=3$):
\be
\begin{array}{c|ccc}
\hbox{Level}& \#\hbox{ of gauge invariant operators}
&\#\hbox{ of independent Gaussian averages}&\#\hbox{ of characters}\\
\hline
&&&\\
1 &1& 1 & 1 \\
2&4&4&4 \\
3 &11& 11 & 11\\
4&43&30&43\\
5&161&61&143\\
6&901&&511\\
7&5579&&1599
\end{array}
\ee
The first and the third columns coincide up to level 4,
because the operators $\chi_{R_1,R_2,R_3}$ are non-zero if and only if $C_{R_1,R_2,R_3}$ are non-zero (see sec.\ref{vanchi}), $C_{R_1,R_2,R_3}$ being the Clebsch-Gordan coefficients for the symmetric group representations , while the number of gauge invariant operators is \cite[eq.(6.19)]{IMMarist}
\be
\#_{\rm g.-inv.ops}=
\sum_{R_1,R_2,R_3\vdash n } C_{R_1R_2R_3}^2
\ee
and non-trivial multiplicities emerge starting from level 5.
By this reason, the numbers in the first column are larger than in the third one at higher levels.

The numbers in the second column are smaller than those in the third one, which is evident
for the following reason: the Gaussian averages are polynomials in $N_i$, with
each $N_i$ entering with the maximal degree of $n$, and all averages are proportional
to $N_1N_2N_3$.
Hence, the linearly independent basis for the Gaussian averages is given,
e.g., by monomials $N_1^{i_1}N_2^{i_2}N_3^{i_3}$, $i_1,i_2,i_3=1,\ldots n$,
and, thus, the number of linearly independent polynomials at level $n$ is restricted by $n^3$:
\be
\#_{\rm lin.indep.G.averages} \leq n^3
\ee
This number grows with $n$ much slower than the number of non-vanishing characters $\chi_{R_1R_2R_3}$,
since the latter are labeled by triples of partitions of $n$.
The number of partitions $P(n)$ at level $n$ grows much faster than $n$,
and so does the number of triples $P(n)^3$.
While one has to subtract from $P(n)^3$ the number of zero Clebsch-Gordan coefficients,
this latter number grows slower\footnote{For instance, the ratio $\xi_n:={ \#_{C_{(n)}\neq 0}\over \#_{C_{(n)}= 0}}$ behaves as a function of $n$ as
\be
{\xi(n)\over n}=0.23, 0.13, 0.14, 0.10, 0.13, 0.11, 0.12, 0.12, 0.13, 0.13\ \ \ \ \ \ \hbox{at}\ \ n=3,\ldots,12
\ee
} than $P(n)^3$ at large $n$. Hence,
\be
 \#_{C\neq 0}=\#_{\chi} \approx \#_{r-{\rm tuples}} \ \ \ \ {\rm as} \ \ \ n\longrightarrow \infty
\ee

Finally,
\be
\#_{\rm lin.indep.G.averages} \leq  \#_{{\rm lin.indep.}\ \chi}
\ee
and there is no equality beginning from $n=4$:
there are linear combinations of characters $\chi_{R_1,R_2,R_3}$
which are non-vanishing by themselves, but have vanishing Gaussian averages.
See (\ref{licochi4}) for the first example.

\bigskip

Thus, we come to the following conclusion:
{\bf
\begin{itemize}
\item The structures with allowed symmetries are the substitutes of characters,
and their averages are products of the corresponding dimensions labeled by Young diagrams
of size $n$.
\item These characters are labeled by triples (generically, $r$-ples) of Young diagrams
of size $n$ but not arbitrary: many triples provide vanishing operators.
\end{itemize}
}

\section{Generalized characters}

Now we give formal definitions of the generalized characters, that is, gauge invariant operators $\chi_{\vec R}\{M,\bar M\}$
as particular linear combinations of ${\cal K}$,
and demonstrate that
they and their Gaussian averages have nice properties,
expected from the character-like quantities.

We also explain why these operators exhaust all which have
non-vanishing averages.

\subsection{Some basic properties of symmetric group characters}

In what follows, we need characters $\psi_R(\gamma)$ of the permutation group $S_n$ (for the theory of permutation groups, see \cite{SG}).
Let us note that the characters effectively depend only on the conjugation class $[\gamma]$
of the permutation: $\psi_R(\gamma) = \psi_R([\gamma])$.
Hereafter, we denote by lower case Greek letters elements of the permutation group,
and, by capital letters the conjugation classes (which are described by Young diagrams).
The number of elements in the conjugacy class $\Delta$ is given by $n!/z_\Delta$,
where $z_\Delta$ is the order of automorphism of the corresponding Young diagram.
We will also need the orthogonality condition
\be\label{orth}
\sum_\gamma \psi_R(\gamma)\psi_Q(\gamma\circ\sigma)=
\sum_\gamma \psi_R(\gamma^{-1})\psi_Q(\gamma\circ\sigma) =
\frac{\psi_R(\sigma)}{d_R}\, \delta_{QR}
\ee
and the value of character on the unit element:
\be
\psi_R(id)=\psi_R([1^{|R|}])=d_R\cdot |R|!
\label{psid}
\ee
In particular, it follows that
\be\label{orth1}
{1\over |R|!}\sum_\gamma \psi_R(\gamma)\psi_Q(\gamma)=
\sum_\Delta {\psi_R(\Delta)\psi_Q(\Delta)\over z_\Delta}=
{\psi_R(id)\over d_R|R|!}\cdot\delta_{RQ}=\delta_{RQ}
\ee
Eq.(\ref{orth}) implies a whole set of identities, and
the simplest one is
\be
\sum_\gamma \psi_R(\sigma_1\circ\gamma\circ\sigma_2\circ\gamma^{-1})=
\frac{\psi_R(\sigma_1)\psi_R(\sigma_2)}{d_R}
\label{orth2}
\ee
The simplest way to prove this identity is to note that the l.h.s. depends only on the conjugacy class of $\sigma_1$ and, hence, one can make a ``Fourier" transform from the conjugacy classes of $\sigma_1$ to the Young diagrams $Q$ given by the kernel $\psi_Q(\sigma_1)$:
\be
\sum_{\sigma_1}\psi_Q(\sigma_1)\sum_\gamma \psi_R(\sigma_1\circ\gamma\circ\sigma_2\circ\gamma^{-1})\stackrel{(\ref{orth})}{=}
\sum_\gamma{\psi_R(\gamma\circ\sigma_2\circ\gamma^{-1})\over d_R}\delta_{RQ}={|R|!\psi_R(\sigma_2)\over d_R}\delta_{RQ}
\ee
Similarly, the same Fourier transform of the r.h.s. of (\ref{orth2}) gives
\be
\sum_{\sigma_1}\psi_Q(\sigma_1)\frac{\psi_R(\sigma_1)\psi_R(\sigma_2)}{d_R}\stackrel{(\ref{orth1})}{=}
{R!\psi_R(\sigma_2)\over d_R}\delta_{RQ}
\ee
The essential thing in this proof is that the Fourier transformation in this case has no kernel.

\subsection{RCM, $r=2$}

One of the possible definitions of the {\bf Schur functions},
depending on the time-variables $p_k$,
expresses them through the characters $\psi_R(\Delta)$:
\be
\chi_R\{p\} = \sum_{\Delta\,\vdash |R|} \frac{ \psi_R(\Delta)}{z_\Delta} p_\Delta\ ,
\label{charvsp}
\ee
where the sum goes over all Young diagrams
$\Delta=\{\delta_1\geq \delta_2\geq\ldots\delta_{l_\Delta}>0\}
= \{1^{m_1},2^{m_2},\ldots\}$
of the same size $|\Delta| \equiv \delta_1+\delta_2+\ldots+\delta_{l }$ as $|R|$,
the symmetry factor is $z_\Delta=\prod_i m_i! \cdot i^{m_i}$,
and $p_\Delta$ is  a monomial
$p_\Delta \equiv p_{\delta_1}p_{\delta_2}\ldots p_{\delta_{l }}$.
The orthogonality of $\psi$
\be
\sum_\Delta \frac{\psi_{R}(\Delta)\psi_{R'}(\Delta)}{z_\Delta} = \delta_{R,R'}
\ee
implies orthogonality of $\chi$
\be
\Big<\hat\chi_R\Big| \chi_{R'}\Big> =  \delta_{R,R'}
\ee
where $\Big<\hat p_\Delta\Big|p_\Delta'\Big> =z_\Delta\delta_{\Delta,\Delta'}$,
i.e. $\hat p_\Delta = z_\Delta \frac{\p}{\p p_{\delta_1}}\ldots\frac{\p}{\p p_{\delta_l}}$.

Gauge invariant operators in RCM are
\be
{\cal K}_{\sigma_1,\sigma_2} = \prod_{p=1}^n M_{a_pb_p}\bar M^{a_{\sigma_1(p)}b_{\sigma_2(p)}}
= {\cal K}_{id,\sigma_1^{-1}\circ \sigma_2}
\ee
Here $\sigma_i$ are elements of the permutation group $S_n$. In fact, ${\cal K}_{id,\sigma}$ depends only on the conjugacy class of $\sigma$,
\be
{\cal K}_{id,\sigma} = P_\Delta
\label{KvsP}
\ee
with $P_k = \Tr (M\bar M)^k$.
We can now introduce a ``Fourier transform"
\be\label{gchr2}
\chi_{R_1,R_2} \equiv   \frac{1}{n!}\sum_{\sigma_1,\sigma_2\in S_n}
\psi_{R_1}(\sigma_1)\psi_{R_2}(\sigma_2){\cal K}_{\sigma_1,\sigma_2}
= \frac{1}{n!}\sum_{\sigma_1,\sigma_2\in S_n}
\psi_{R_1}(\sigma_1)\psi_{R_2}(\sigma_2){\cal K}_{id, \sigma_1^{-1}\circ\sigma_2}
\ \stackrel{\sigma_2\to\sigma_1\circ\sigma_2}{=} \nn \\
= \frac{1}{n!}\sum_{\sigma_1,\sigma_2\in S_n}
\psi_{R_1}(\sigma_1)\psi_{R_2}(\sigma_1\circ\sigma_2){\cal K}_{id,\sigma_2}
\stackrel{(\ref{orth})}{=} \frac{1}{n!}\cdot\frac{\delta_{R_1,R_2}}{d_{R_1}}
\sum_{\sigma_2} \psi_{R_2}(\sigma_2){\cal K}_{id,\sigma_2}
= \frac{\delta_{R_1,R_2}}{d_{R_1}} \chi_{R_2}\{P\}
\label{chivsodchi}
\ee
where at the last stage we used the fact that ${\cal K}_\sigma$ depends only on the
conjugacy class $\Delta$ of $\sigma$ so that
\be
\frac{1}{n!}  \sum_\sigma \psi_{R }(\sigma ){\cal K}_{id,\sigma }
= \sum_{\Delta\vdash n} \frac{\psi_R(\Delta)}{z_\Delta}{\cal K}_{id,\sigma }
\ \stackrel{(\ref{KvsP})}{=} \
\sum_{\Delta\vdash n} \frac{\psi_R(\Delta)}{z_\Delta}P_\Delta
\ \stackrel{(\ref{charvsp})}{=} \ \chi_R\{P\}
\label{odchivsK}
\ee
The transformation (\ref{gchr2}) is not invertible, if considered as
a map from the space of functions $F_{\sigma_1,\sigma_2}$ to $\chi_{R_1,R_2}$,
because the space of all possible $R$ at fixed $|R|=n$ and that of all possible $\sigma$ in $S_n$ have different dimensions: the number of $\sigma$, which is equal to $n!$, is larger than the number of Young diagrams that enumerate the conjugacy classes of $\sigma$.
However, the actual operators ${\cal K}_{\sigma_1,\sigma_2}$ actually depend only on the product $\sigma^{-1}\circ\sigma_2$, i.e. only on its conjugacy class.
Thus, they are in one-to-one correspondence with characters, and are diagonal, $\chi_{R_1,R_2} \sim \delta_{R_1,R_2}$.

To summarize, our $\chi_{R_1,R_2}$ is essentially the character:
\be
\boxed{
\chi_{R_1,R_2} = \frac{\delta_{R_1,R_2}}{d_{R_1}}\ \chi_{R_1}\{P\}
}
\ee
and its Gaussian average is  fully symmetric in $R_1$ and $R_2$:
\be
\boxed{
\Big<\chi_{R_1,R_2}\Big> =  \delta_{R_1,R_2}\,\frac{D_{R_1}(N_1)D_{R_2}(N_2)}{d_{R_1}d_{R_2}}
}
\ee
An advantage of this redefinition is that now it can be straightforwardly generalized
beyond $r=2$, i.e. from matrix to tensor models.

\subsection{Aristotelian model, $r=3$}

Consider the Aristotelian $r=3$ case. Denote
\be\label{ar}
{\cal K}_{\sigma_1\sigma_2\sigma_3} =
\prod_{p=1}^n M_{a_pb_pc_p}\bar M^{a_{\sigma_1(p)}b_{\sigma_2(p)}c_{\sigma_3(p)}}
\ee
(as usual, summation is assumed over all $a_p$, $b_p$ and $c_p$).
Now, one can introduce the quantity
\be
\boxed{
\chi_{R_1R_2R_3}:=\frac{1}{n!}
\sum_{\{\sigma_i\}\in S_n}\psi_{R_1}(\sigma_1)\psi_{R_2}(\sigma_2)\psi_{R_3}(\sigma_3)\cdot
{\cal K}_{\sigma_1\sigma_2\sigma_3}
}
\label{charArist}
\ee
which generalizes the notion of character.

Unlike in RCM with $r=2$,
already in the Aristotelian case of $r=3$
the sum can not be reduced to summation over the conjugation classes only.
What remains is the sum
over the elements of (the orbit of) the permutation group:
in this case, the counterpart of (\ref{gchr2}) is
\be\label{gch3}
\chi_{R_1R_2R_3}=\frac{1}{n!}
\!\!\!\!\!\!\!\!\!\!\!\!\!\!\!\!\!\!\!
\sum_{\ \ \ \ \ \ \ \ \ _{\{\sigma_1,\sigma_2,\sigma_3\}\in S_n}}
\!\!\!\!\!\!\!\!\!\!\!\!\!\!\!\!\!\!
\psi_{R_1}(\sigma_1)\psi_{R_2}(\sigma_2) \psi_{R_3}(\sigma_3)
\,{\cal K}_{\sigma_1\sigma_2\sigma_3}
= \frac{1}{n!}
\!\!\!\!\!\!\!\!\!\!\!\!\!\!\!\!\!\!\!
\sum_{\ \ \ \ \ \ \ \ \ _{\{\sigma_1,\sigma_2,\sigma_3\}\in S_n}}
\!\!\!\!\!\!\!\!\!\!\!\!\!\!\!\!\!\!
 \psi_{R_1}(\sigma_1)\psi_{R_2}(\sigma_1\circ\sigma_2) \psi_{R_3}(\sigma_1\circ\sigma_3)
\,{\cal K}_{id, \sigma_2,\sigma_3}
\label{chidef3}
\ee
where we used the property
${\cal K}_{\sigma_1\sigma_2\sigma_3}
={\cal K}_{id,\sigma_1^{-1}\circ\sigma_2,\sigma_1^{-1}\circ \sigma_3}$
which follows from the definition (\ref{ar}).
In fact, there is a larger symmetry: one can preserve the form ${\cal K}_{id,\sigma_2,\sigma_3}$, while
making the conjugation $\sigma_2\to\gamma^{-1}\circ\sigma_2\circ\gamma$,
$\sigma_3\to\gamma^{-1}\circ\sigma_3\circ\gamma$. We will ignore this fact in what follows.
In any case, the transformation (\ref{charArist}) is not one-to-one on any reasonable space,
both on the left and on the right.
This means that our $\chi_{_{R_1R_2R_3}}$ is both insufficient to describe the entire space of gauge invariant operators
and redundant to describe the space of of gauge invariant operators with non-vanishing Gaussian averages. It instead serves as {\it a generalized character}
$\chi_{\vec R}$, which is simple by itself and closed under simple operations.

In fact, as we saw in examples in the previous sections, there is a much stronger
statements: that $\Big<\chi_{R_1,R_2,R_3}\Big>$ form  a (redundant) basis in the space
of all Gaussian averages.
Taking an explicit expression for the Gaussian average of ${\cal K}_{\sigma_1\sigma_2\sigma_3}$
from \cite{rainbow,MMtenmod}, we obtain:
\be\label{Ga2}
\Big< {\cal K}_{\sigma_1\sigma_2\sigma_3}\Big>=
\sum_{{\{Q_i\}\vdash n}\atop{\gamma\in S_n}}
\prod_{i=1}^3D_{Q_i}(N_i)\psi_{Q_i}(\gamma\circ\sigma_i)
\ee
and, using the orthogonality condition (\ref{orth}),
one immediately obtains (see also \cite[eq.(79)]{Ramg2})
\be\boxed{
\Big<\chi_{R_1R_2R_3}\Big>
= C_{R_1R_2R_3}\cdot
{D_{R_1}(N_1)\over d_{R_1}}{D_{R_2}(N_2)\over d_{R_2}} {D_{R_3}(N_3)\over d_{R_3}}}
\label{avecharArist}
\ee
where
\be
C_{R_1R_2R_3}:={1\over n!}\sum_{\gamma\in S_n}
{\psi_{R_1}(\gamma)\psi_{R_2}(\gamma)\psi_{R_3}(\gamma)}=
\sum_{\Delta\vdash n}{\psi_{R_1}(\Delta)\psi_{R_2}(\Delta)\psi_{R_3}(\Delta)\over z_\Delta}
\label{C3}
\ee
are the Clebsch-Gordan coefficients, which vanish in the case of forbidden symmetries.

\subsection{
Vanishing of $\chi_{\vec R}$ with forbidden symmetries
\label{vanchi}}

Let us prove that vanishing the Clebsch-Gordan coefficients implies not only vanishing of the Gaussian averages, but also the generalized characters $\chi_{\vec R}$ themselves.

The coefficients
\be
C^{\sigma_2,\sigma_3}_{R_1,R_2,R_3}:= \frac{1}{n!}
\sum_{  _{\gamma \in S_n}}
 \psi_{R_1}(\gamma)\psi_{R_2}(\gamma\circ\sigma_2) \psi_{R_3}(\gamma\circ\sigma_3)
 \label{Cs3}
\ee
in the definition  (\ref{chidef3}) of $\chi_{R_1R_2R_3}$
satisfy the orthogonality condition
\be
\sum_{\sigma_1,\sigma_2}C^{\sigma_1\sigma_2}_{R_1R_2R_3}C^{\sigma_1\sigma_2}_{Q_1Q_2Q_3}=
{C_{R_1R_2R_3}\over d_{R_1}d_{R_2}d_{R_3}}\delta_{R_1Q_1}\delta_{R_2Q_2}\delta_{R_3Q_3}
\ee
This means that
\be
\sum_{\sigma_1,\sigma_2}\Big(C^{\sigma_1\sigma_2}_{R_1R_2R_3}\Big)^2=
{C_{R_1R_2R_3}\over d_{R_1}d_{R_2}d_{R_3}}
\ee
and, since at the l.h.s. we have a sum of squares of rational real-valued quantities,
we get as an immediate corollary
\be\label{C=0}
\boxed{
C_{R_1R_2R_3}=0 \ \ \Longrightarrow \ \
{\rm all} \ \ C^{\sigma_1\sigma_2}_{R_1R_2R_3}=0 \ \ \stackrel{(\ref{chidef3})}{\Longrightarrow} \ \ \chi_{R_1R_2R_3}=0
}
\ee
Thus we {\it proved} that vanishing of the Clebsh-Gordan coefficient $C_{R_1R_2R_3}=0$
implies {\it identical vanishing} of the corresponding generalized character $\chi_{R_1R_2R_3}$.

\subsection{Generalized characters as a basis for operators with non-vanishing Gaussian averages\label{nvG}}

Using the identity (\ref{C=0}), we can now prove that the generalized characters form a basis in the space of all operators with non-vanishing Gaussian averages. To this end, it is enough to prove that any Gaussian average can be written as a linear combination of the generalized characters. Indeed, any Gaussian average is given by formula (\ref{Ga2}). In fact, as we already discussed, it is enough to consider the average of ${\cal K}_{id,\sigma_2\sigma_3}$,
\be
\Big< {\cal K}_{id,\sigma_2\sigma_3}\Big>=
\sum_{{\{R_i\}\vdash n}\atop{\gamma\in S_n}}C^{\sigma_2\sigma_3}_{R_1R_2R_3}
\prod_{i=1}^3D_{R_i}(N_i)
\ee
In the sum, contribute only $R_i$ such that $C^{\sigma_1\sigma_2}_{R_1R_2R_3}\ne 0$, which, as follows from (\ref{C=0}), simultaneously implies $C_{R_1R_2R_3}\ne 0$. In this case, one obtains from (\ref{avecharArist}) that
\be
\Big< {\cal K}_{id,\sigma_2\sigma_3}\Big>=
\sum_{{\{R_i\}\vdash n}\atop{\gamma\in S_n}}C^{\sigma_2\sigma_3}_{R_1R_2R_3}
{d_{R_1}d_{R_2}d_{R_3}\over C_{R_1R_2R_3}}\Big<\chi_{R_1R_2R_3}\Big>
\ee
i.e. any Gaussian average can be, indeed, written as a linear combination of the generalized characters. Moreover,
since the coefficients in this combination does not depend on $N_i$'s, and the latter enter only through the generalized characters,
this implies that the generalized characters form a basis in the space of all gauge invariant operators with non-vanishing Gaussian averages.

Let us point out that we understand by the space of all gauge invariant operators with non-vanishing Gaussian averages the space with two operators equivalent if their difference is an operator with vanishing Gaussian average. In fact, the Gaussian averages of the operators ${\cal K}$, (\ref{ar}) associated with concrete permutations are non-vanishing, which follows both from the explicit examples of \cite[eqs.(7.9),(7.15), s.7.4.1, etc.]{IMMarist} and from associating the large $N$ limit of these Gaussian averages with (non-vanishing) Feynman diagrams in the matrix model \cite{FDTM}.
However, the Gaussian averages of these operators are subject to vanishing linear combinations, and all of them are spanned by the generalized characters up to operators with vanishing Gaussian averages.

\subsection{Background metrics}

For the action $M_{abc}\bar M^{\bar a\bar b\bar c}A^{a}_{\bar a}B^{b}_{\bar b}C^{c}_{\bar c}$,
we get a generalization of the ``cut" relation (\ref{avecharArist})
\be
\Big<\chi_{R_1R_2R_3} \Big> = C_{R_1R_2R_3} \frac{\chi_{R_1}[ A] \chi_{R_2}[B]\chi_{R_3}[C]}
{d_{R_1}d_{R_2}d_{R_3}}
\ee
Another  matrix model relation (``join"),
\be
\Big<\chi_{R}[MC]\chi_{R'}[\bar MD]\Big> = \frac{\chi_R[CD]}{d_R}\cdot\delta_{R,R'}
\ee
seems not to have a direct generalization to $r>2$.

\subsection{Other rainbow models, arbitrary $r$}

Similarly, for arbitrary $r$, we define
\be
\chi_{R_1,\ldots,R_r} = \frac{1}{n!}\sum_{\sigma_1,\ldots,\sigma_r\in S_n}
\psi_{R_1}(\sigma_1)\ldots\psi_{R_r}(\sigma_r)\,{\cal K}_{\sigma_1,\ldots,\sigma_r}
\ee
where, as a generalization of (\ref{ar}),
\be
\label{arr}
{\cal K}_{\vec\sigma} = {\cal K}_{\sigma_1 \ldots \sigma_r} =
\prod_{p=1}^n M_{a_p^{(1)}\ldots a_p^{(r)}}
\bar M^{a^{(1)}_{\sigma_1(p)}\ldots a^{(r)}_{\sigma_r(p)}}
= \prod_{p=1} M_{\vec a_p}\bar M^{\vec a_{\vec\sigma(p)}}
\ee
The Gaussian averages are
\be
\sum_{\vec \mu}  \left(\Big<{\cal K}_{\vec \mu}\Big>\cdot \prod_{i=1}^r \psi_{R_i}(\mu_i)\right)
=C_{\vec R} \cdot\prod_{i=1}^r   \frac{D_{R_i}(N_i)}{d_{R_i}}
\ee
where
\be\label{CG}
C_{\vec R}:={1\over n!}\sum_{\gamma\in S_n}\prod_{i=1}^r\psi_{R_i}(\gamma)
=\sum_{\Delta\vdash n}{\prod_{i=1}^r \psi_{R_i}(\Delta)\over z_\Delta}
\ee
These coefficients can be also expressed through the Clebsch-Gordan coefficients (\ref{C3}):
\be
C_{\vec R}=\sum_{\{Q_i\}} C_{R_1R_2Q_1}C_{Q_1R_3Q_2}C_{Q_2R_4Q_3}\ldots C_{Q_{r-3}R_{r-1}R_r}
\ee

\section{Orthogonality of generalized characters}

\subsection{The problem}

The first essential property of characters is orthogonality.
Conventional characters satisfy the orthogonality condition \cite{Mac}:
\be
\boxed{
\Big<\chi_R\Big|\chi_{R'}\Big>=\delta_{RR'}
}
\ee
where the scalar product is explicitly given by
\be\label{sp1}
\Big<\chi_R\Big|\chi_{R'}\Big>:=\chi_R(k\partial_k)\chi_{R'}(p_k)\Big|_{p_k=0}
\ee
Indeed, parameterizing the Young diagram $\Delta$ by the numbers $m_i$
which counts the number of lines of the same length $i$ in the diagram
(i.e., for instance, for $\Delta=[4,2,2,1]$, $m_1=1$, $m_2=2$, $m_3=0$, and $m_4=1$)
and taking into account that, in these terms, $z_\Delta=\prod_{i:m_i\ne 0} i^{m_i}m_i!$
with the product over all $i$ with non-zero $m_i$, one obtains
\be
\left.\chi_R\Big\{k\frac{\partial}{\p p_k}\Big\}\chi_{R'}\{p_k\}\right|_{p_k=0}
=\sum_{\Delta,\Delta'}{\psi_R(\Delta)\over z_\Delta}{\psi_{R'}(\Delta')\over z_{\Delta'}}
\prod_j \left(j{\partial\over\partial p_j}\right)^{m_j}
\prod_i p_i^{m'_j}\ \longrightarrow \ \ \
\nn \\ \stackrel{p_k=0}{\longrightarrow} \
\sum_{\Delta,\Delta'}{\psi_R(\Delta)\over z_\Delta}{\psi_{R'}(\Delta')\over z_{\Delta'}}
 \prod_{i:m_i\ne 0} i^{m_i}m_i!\delta_{m_i,m'_i}=
\sum_{\Delta,\Delta'}{\psi_R(\Delta)\over z_\Delta}{\psi_{R'}(\Delta')
\over z_{\Delta'}}z_\Delta\delta_{\Delta\Delta'}=
\sum_{\Delta}{\psi_R(\Delta)\psi_{R'}(\Delta)\over z_{\Delta}}=\delta_{RR'}
\ee
where, in the last equality, we have used the orthogonality relation (\ref{orth1}).
Thus, orthogonality of $\chi$ is essentially reduced to orthogonality of $\psi$.

However, above manipulation heavily depends on existence of $p$-variables,
moreover, involve derivatives with respect to $p$.
In this form, it has low chances for tensorial generalization.
Fortunately, in \cite{MMcharMAMOreview} the first step was done towards
elimination of $p$-variables and reformulation of the theory of $W$-operators
directly in terms of $\chi$-variables.
In what follows, we introduce an even more powerful and straightforward formalism,
which remains to be properly understood, but is already sufficient for
tensor model applications.

It involves two ideas.
First, we substitute the monomials $p_\Delta$
by independent {\it linear} variables $\xi_\Delta$.
Second, we extend this set of variables to $\xi_\sigma$
depending on permutations $\sigma$ rather than on their
conjugation classes.
Then, the dual characters, which were obtained by the substitution
$p_k \longrightarrow k\p/\p p_k$, where the coefficient $k$
somehow ``remembers" about non-linearity in $p$,
are described by a linear substitution $\xi_\sigma\longrightarrow
\p/\p\xi_\sigma$ without any $\sigma$-dependent coefficients.
This is already nice, but most important, this formalism
continues to work for tensors.
Moreover, the apparent redundancy of the $\xi_\sigma$-variables
turns out to be exactly what is needed to capture the set of
${\cal K}_{\vec\sigma}$ with non-vanishing Gaussian averages.

\subsection{RCM, $r=2$, through ordinary characters}

Following this plan, we substitute
\be
\chi_R\{p\} = \sum_{\Delta\vdash |R|} \frac{\psi_R(\Delta)}{z_\Delta}p_\Delta
\ \longrightarrow \
\tchi_R(\xi) = \sum_{\sigma\in S_n} \psi_R(\sigma)\cdot \xi_\sigma
\label{tochivsxi2}
\ee
where $n=|R|$ and $\psi_R(\sigma):=\psi_R([\sigma])$
depends only on the conjugation class
$\Delta=[\sigma]$ of the permutation $\sigma$.
Since there are $\frac{n!}{z_{\Delta}}$
different permutations in this class,
we read from these relations and (\ref{gchr2})  that
\be
\sum_{\sigma\in \Delta} \xi_\sigma = \frac{1}{n!}\sum_{\sigma\in \Delta}{\cal K}_{id,\sigma}
= \frac{1}{z_\Delta}\, P_{\Delta}
\ee

Then, since
\be
\Big<p_\Delta|p_{\Delta'}\Big> =
\Big<p_1^{m_1}p_2^{m_2}p_3^{m_3}\ldots\Big|p_1^{m_1'}p_2^{m_2'}p_3^{m_3'}\ldots\Big>
= \left.\p_{p_1}^{m_1}(2\p_{p_2})^{m_2}(3\p_{p_3})^{m_3}\ldots
\ p_1^{m_1'}p_2^{m_2'}p_3^{m_3'}\ldots\ \right|_{p=0}
= z_\Delta \,\delta_{\Delta,\Delta'}
\ee
it is natural to postulate a new scalar product for linear functions
of $\xi$-variables:
\be\label{sp2}
\BBigl\xi_{\sigma}\Big|\xi_{\sigma'}\BBigr =\frac{\delta_{\sigma\sigma'}}{n!}
\ee
Indeed, in this case
\be
\BBigl P_\Delta\Big|P_{\Delta'}\BBigr = z_\Delta z_{\Delta'}
\sum_{\sigma\in \Delta}\sum_{\sigma'\in \Delta'}
\BBigl\xi_{\sigma}\Big|\xi_{\sigma'}\BBigr
= z_\Delta z_{\Delta'}\,\delta_{\Delta,\Delta'}  \sum_{\sigma\in \Delta} \frac{1}{|\sigma|!}
= z_\Delta\,\delta_{\Delta,\Delta'}
\ee
Then we immediately obtain
\be
\BBigl \tchi_R\Big|\tchi_{R'}\BBigr =
\sum_{\sigma,\sigma'\in S_n}
\psi_R(\sigma)\psi_{R'}(\sigma')
\BBigl\xi_{\sigma}\Big|\xi_{\sigma'}\BBigr
= \frac{1}{n!}\sum_{\sigma \in S_n}
\psi_R(\sigma)\psi_{R'}(\sigma)
\stackrel{(\ref{orth1})}{=}  \delta_{R,R'}
\ee

\subsection{RCM, $r=2$, through bi-characters}

If we keep in mind our goal of generalization from matrix to tensor models,
we need now to rewrite the previous section in terms of the generalizable
object, the bi-character $\chi_{R_1,R_2}$ instead of the ordinary one $\chi_R$.
The natural counterpart of (\ref{tochivsxi2}) would be
\be
\tchi_{R_1,R_2}(\eta):=\sum_{\sigma_1,\sigma_2\in S_n} \psi_{R_1}(\sigma_1)
\psi_{R_2}(\sigma_2)\cdot \eta_{\sigma_1,\sigma_2}
\ee
with new indeterminants $\eta$.
However, this degree of redundancy is too much even for tensor model
generalizations.
We can safely demand that $\eta$ like ${\cal K}$ operators
is invariant under simultaneous conjugations,
$\eta_{\sigma_1,\sigma_2} = \eta_{\sigma\circ\sigma_1\circ\sigma^{-1},
\sigma\circ\sigma_2\circ\sigma^{-1}}$, what allows one to substitute
$\eta_{\sigma_1,\sigma_2} = \eta_{id,\sigma_1^{-1}\circ\sigma_2}$,
which is essentially the $\xi$-variables.

Namely, using (\ref{gchr2}) and  (\ref{tochivsxi2}),  introduce (hereafter, we use the notation $\tilde\chi$ for the generalized character in $\xi$-variables)
\be
\tchi_{R_1R_2}(\xi)={\delta_{R_1R_2}\over d_{R_1}}
\sum_{\sigma\in S_n}\psi_{R_2}(\sigma)\cdot{\xi}_{\sigma}
\label{chivsxi}
\ee
where $|R_1|=|R_2|=n$ and the scalar product is (\ref{sp2}).
Then
\be
\boxed{
\BBigl\tchi_{R_1R_2}\Big|\tchi_{R'_1R'_2}\BBigr
={\delta_{R_1R_2}\delta_{R'_1R'_2}\over d_{R_1}d_{R'_1}}
\cdot\frac{1}{n!}\,\sum_{\gamma\in S_n}  \psi_{R_2}(\gamma)\psi_{R'_2}(\gamma)
\stackrel{(\ref{orth1})}{=}
  {\delta_{R_1R_2}\delta_{R'_1R'_2}\delta_{R_1R_1'}\over d_{R_1}^2}
}
\ee

One can also realize the scalar product (\ref{sp2}) similarly to (\ref{sp1}):
\be
\left.\BBigl\tchi_{RR}\Big|\tchi_{R'R'}\BBigr:= \frac{1}{|R|!\cdot d_R d_{R'}}\cdot
\tilde\chi_R\Big({\partial\over\partial \xi }\Big)\tilde\chi_{R'}(\xi )\right|_{\xi=0}
\ee
where $\tilde\chi_R(\xi)$ is defined in (\ref{tochivsxi2}).

\bigskip

For {\bf example}, at level $n=2$
\be
\tilde\chi_{[2],[2]} \ \stackrel{(\ref{chivsodchi})}{=}\
{\cal K}_{id,id}+{\cal K}_{id,(12)} = P_1^2+P_2 = 2\tilde\chi_{[2]}
\ \stackrel{(\ref{chivsxi})}{=}\  2(\xi_{id}+\xi_{(12)}) \nn \\
\tilde\chi_{[1,1],[1,1]} \ \stackrel{(\ref{chivsodchi})}{=}\
{\cal K}_{id,id}-{\cal K}_{id,(12)} = P_1^2-P_2 = 2\tilde\chi_{[1,1]}
\ \stackrel{(\ref{chivsxi})}{=}\ 2( \xi_{id}-\xi_{(12)})
\ee
and
\be
\BBigl\tchi_{[1,1],[1,1]}\Big|\tchi_{[1,1],[1,1]}\BBigr
= 4\BBigl\tchi_{[1,1] }\Big|\tchi_{[1,1] }\BBigr
= \frac{2^2}{2!}\,\Big(\frac{\p}{\p {\xi_{id}}}+\frac{\p}{\p {\xi_{(12)}}}\Big)
\Big(\xi_{(id)}+\xi_{(12)}\Big)=4 = \frac{1}{d_{[1,1]}^2} \nn \\
\BBigl\tchi_{[2],[2]}\Big|\tchi_{[1,1],[1,1]}\BBigr
=4\BBigl\tchi_{[2] }\Big|\tchi_{[1,1] }\BBigr
= \frac{2^2}{2!}\,\Big(\frac{\p}{\p {\xi_{id}}}-\frac{\p}{\p {\xi_{(12)}}}\Big)
\Big(\xi_{id}+\xi_{(12)}\Big) = 0
\nn \\
\BBigl\tchi_{[1,1],[1,1]}\Big|\tchi_{[2],[2]}\Big>
=4\BBigl\tchi_{[1,1] }\Big|\tchi_{[2] }\Big>
= \frac{2^2}{2!}\,\Big(\frac{\p}{\p {\xi_{id}}}+\frac{\p}{\p {\xi_{(12)}}}\Big)
\Big(\xi_{id}-\xi_{(12)}\Big) = 0
\nn \\
\BBigl\tchi_{[2],[2]}\Big|\tchi_{[2],[2]}\BBigr
= 4\BBigl\tchi_{[2] }\Big|\tchi_{[2] }\BBigr
= \frac{2^2}{2!}\,\Big(\frac{\p}{\p {\xi_{id}}}-\frac{\p}{\p {\xi_{(12)}}}\Big)
\Big(\xi_{id}-\xi_{(12)}\Big)=4 = \frac{1}{d_{[2]}^2}
\ee

The difference between permutations and their conjugation classes
first shows up at level $n=3$:
\be
\tchi_{[3],[3]} \ \stackrel{(\ref{chivsodchi})}{=}\
&\!\!\!\!\!\!\!\!\!\!\!\!\!\frac{6}{3!}
\Big({\cal K}_{id,id}+ {\cal K}_{id,(12)}+{\cal K}_{id,(13)}+{\cal K}_{id,(23)}
+ {\cal K}_{id,(123)}+{\cal K}_{id,(132)}\Big) = \nn \\
&=  P_1^3+3P_2P_1+2P_3  = 6\tchi_{[3]}
\ \stackrel{(\ref{chivsxi})}{=}\
6\Big(\xi_{id}+\xi_{(12)}+\xi_{(13)}+\xi_{(23)}
+ \xi_{(123)}+\xi_{(132)}\Big) \nn \\
\tchi_{[2,1],[2,1]} \ \stackrel{(\ref{chivsodchi})}{=}\
&\frac{3}{3!}
\Big(2{\cal K}_{id,id} -{\cal K}_{id,(123)}-{\cal K}_{id,(132)}\Big)
%= \nn \\
=\frac{1}{2}\Big(2P_1^3 -2P_3\Big) = 3\tchi_{[2,1]}
\ \stackrel{(\ref{chivsxi})}{=}\  3\Big(2\xi_{id}
- \xi_{(123)}-\xi_{(132)}\Big) \!\!\!\!\!\!\!\!\!\!\!\!\!\!\!\!\!\!\!\!\!\!\!\!\!\!\nn \\
\!\!\!\!\!\!\!\!\!\!\!\!\!
\tchi_{[1,1,1],[1,1,1]} \ \stackrel{(\ref{chivsodchi})}{=}\
&\!\!\!\!\!\!\!\!\!\!\!\!\!\frac{6}{3!}
\Big({\cal K}_{id,id}- {\cal K}_{id,(12)}-{\cal K}_{id,(13)}-{\cal K}_{id,(23)}
+ {\cal K}_{id,(123)}+{\cal K}_{id,(132)}\Big)
= \nn \\
&= P_1^3-3P_2P_1+2P_3  = 6\tchi_{[1,1,1]}
\ \stackrel{(\ref{chivsxi})}{=}\  6\Big(\xi_{id}-\xi_{(12)}-\xi_{(13)}-\xi_{(23)}
+ \xi_{(123)}+\xi_{(132)}\Big)
\ee
so that
{\footnotesize
\be
\BBigl\tchi_{[1,1,1],[1,1,1]}\Big|\tchi_{[1,1,1],[1,1,1]}\BBigr
= 6^2\BBigl\tchi_{[1,1,1] }\Big|\tchi_{[1,1,1] }\BBigr  = \nn \\
= \frac{6^2}{3!}\,\Big(\frac{\p}{\p {\xi_{id}}}+\frac{\p}{\p {\xi_{(12)}}}
+\frac{\p}{\p {\xi_{(13)}}}+\frac{\p}{\p {\xi_{(23)}}}
+\frac{\p}{\p {\xi_{(123)}}}+\frac{\p}{\p {\xi_{(132)}}}\Big)
\Big(\xi_{id}+\xi_{(12)}+\xi_{(13)}+\xi_{(23)}+\xi_{(123)}+\xi_{(132)}\Big)
=36 = \frac{1}{d_{[1,1,1]}^2}
\nn \\
\BBigl\tchi_{[2,1],[2,1]}\Big|\tchi_{[2,1],[2,1]}\BBigr
= 3^2\BBigl\tchi_{[2,1] }\Big|\tchi_{[2,1] }\BBigr
%= \nn \\
= \frac{3^2}{3!}\,\Big(2\frac{\p}{\p {\xi_{id}}}
-\frac{\p}{\p {\xi_{(123)}}}-\frac{\p}{\p {\xi_{(132)}}}\Big)
\Big(2\xi_{id}-\xi_{(123)}-\xi_{(132)}\Big)
=9 = \frac{1}{d_{[2,1]}^2}
\nn
\ee
}

\vspace{-0.4cm}
\noindent
and so on.

To conclude this section, we emphasize once again that
$\tchi$ are just {\it linear} functions of $\xi$-variables,
and conjugate to $\xi$ are just $\xi$-derivatives,
with nothing like factors $k$ in the conventional formalism with  $p_k$-derivatives.

\subsection{Aristotelian model, $r=3$}

Now we consider the case of the Aristotelian model.
In this case the character depends not only on the sums over the conjugation classes,
hence, we need the full set of $\xi$-variables.
Thus, the generalized character is
\be
\tchi_{R_1R_2R_3}:= \frac{1}{n!}
\sum_{\{\sigma_i\}\in S_n}
\psi_{R_1}(\sigma_1)\psi_{R_2}(\sigma_2)\psi_{R_3}(\sigma_3)\cdot \eta_{\sigma_1\sigma_2\sigma_3}
\label{chivseta3}
\ee
where $\eta_{\sigma_1\sigma_2\sigma_3}$ are indeterminants with the property
$\eta_{\sigma_1\sigma_2\sigma_3}=\eta_{id,\sigma_1^{-1}\circ\sigma_2,\sigma_1^{-1}\circ \sigma_3}$.
Hence, similarly to (\ref{gch3}), one can consider
\be
\tchi_{R_1R_2R_3}:=\sum_{\gamma,\sigma_1,\sigma_2\in S_n}
\psi_{R_1}(\gamma)\psi_{R_2}(\gamma\circ \sigma_1) \psi_{R_3}(\gamma\circ\sigma_2)
\cdot \xi_{\sigma_1\sigma_2}
\label{chivsxi3}
\ee
with free\footnote{\label{var}This is an essential point:
we use {\it more} variables than  necessary, there is an additional remaining symmetry
$\xi_{\sigma_1\sigma_2}=\xi_{\gamma^{-1}\circ\sigma_1\circ\gamma,\gamma^{-1}
\circ\sigma_2\circ\gamma}$, see \cite{IMMarist}
(in terms of that paper, we are working here in the {\bf RG}-gauge).
This means that the generalized characters effectively depend on less variables:
they depend only on some combinations of $\xi_{\sigma_1\sigma_2}$.
They can be chosen, for instance, as sums over conjugacy classes of $\sigma_2$,
similarly to what we did in the case of $r=2$.
However, we ignore this subtlety here.}
indeterminants $\xi_{\sigma_1\sigma_2}$.
Then, if we require that
\be
\BBigl\xi_{\sigma_1\sigma_2}\Big|\xi_{\sigma'_1\sigma'_2}\BBigr
=\frac{\delta_{\sigma_1\sigma'_1} \delta_{\sigma_2\sigma'_2}}{n!}
\label{xixi30}
\ee
and apply (\ref{orth}), we get:
\be
\BBigl \tchi_{R_1R_2R_3}\Big|\tchi_{R'_1R'_2R'_3}\BBigr
=\frac{1}{n!} \sum_{\gamma,\gamma',\sigma_1,\sigma_2\in S_n}
\psi_{R_1}(\gamma)\psi_{R_2}(\gamma\circ\sigma_1)\psi_{R_3}(\gamma\circ\sigma_2)
\psi_{R'_1}(\gamma')\psi_{R'_2}(\gamma'\circ\sigma_1)\psi_{R'_3}(\gamma'\circ\sigma_2)
=\!\!\!\!\!\!\!\!\!\!\nonumber\\
={\delta_{R_2R'_2}\delta_{R_3R'_3}\over d_{R_2}d_{R_3}}
\cdot\frac{1}{n!}\,
\sum_{\gamma,\gamma'\in S_n} \psi_{R_1}(\gamma)\psi_{R_2}(\gamma'\circ\gamma^{-1})
\psi_{R_3}(\gamma'\circ\gamma^{-1}) \psi_{R'_1}(\gamma')
\stackrel{\gamma'\to\gamma'\circ\gamma}{=}\nn\\
= {\delta_{R_2R'_2}\delta_{R_3R'_3}\over d_{R_2}d_{R_3}}
\cdot\frac{1}{n!}
\sum_{\gamma,\gamma'\in S_n} \psi_{R_1}(\gamma)\psi_{R_2}(\gamma')\psi_{R_3}(\gamma')
\psi_{R'_1}(\gamma'\circ\gamma)
\stackrel{(\ref{orth})}{=}
\nn \\
=   {\delta_{R_1R'_1}\delta_{R_2R'_2}\delta_{R_3R'_3}\over d_{R_1}d_{R_2}d_{R_3}}
\cdot\frac{1}{n!}\sum_{ \gamma'\in S_n} \psi_{R_1}(\gamma')\psi_{R_2}(\gamma')\psi_{R_3}(\gamma')
= {C_{R_1R_2R_3}\over d_{R_1}d_{R_2}d_{R_3}}\
\delta_{R_1R'_1}\delta_{R_2R'_2}\delta_{R_3R'_3}
\ee
where we used the definition of $C$ in (\ref{C3}).
Finally, with the definitions (\ref{chivsxi3}) and (\ref{xixi30}),
\be
\boxed{
\BBigl \tchi_{R_1R_2R_3}\Big|\tchi_{R'_1R'_2R'_3}\BBigr
= {C_{R_1R_2R_3}\over d_{R_1}d_{R_2}d_{R_3}}\
\delta_{R_1R'_1}\delta_{R_2R'_2}\delta_{R_3R'_3}
}
\label{chichi3}
\ee
Comparing with (\ref{charArist}) we see that $\eta$ in (\ref{chivseta3}) and thus
$\xi$ in (\ref{chivsxi3}) just substitute ${\cal K}$, while (\ref{xixi30}) {\it defines}
a scalar product in the space of the relevant operators ${\cal K}$,
which makes our $\chi_{\vec R}$ orthogonal.

\subsection{Other rainbow models, arbitrary $r$}

In complete analogy with the $r=3$ case, one can consider an arbitrary $r$.
Then we define
\be
\tchi_{\vec R}:=\sum_{\gamma,\{\sigma_j\}\in S_n}
\psi_{R_1}(\gamma)\prod_{j=1}^{r-1}\psi_{R_{j+1}}(\gamma\circ \sigma_j)\cdot \xi_{\vec \sigma}
\ee
require that
\be
\BBigl \xi_{\vec \sigma}\Big|\xi_{\vec \sigma'}\BBigr
=\frac{1}{n!}\prod_{i=1}^{r-1}\delta_{\sigma_i\sigma'_i}
\ee
and obtain
\be
\boxed{
\BBigl \tchi_{\vec R}\Big|\tchi_{\vec R'}\BBigr
= {C_{\vec R}\over \prod_{i=1}^rd_{R_i}}\ \prod_{i=1}^r\delta_{R_iR'_i}
}
\label{chichir}
\ee

\bigskip

Despite somewhat formal introduction of {\it linear} $\xi$-variables and associated
scalar product, this allows us to make the next step:
to introduce generalized cut-and-join operators $\hat W$,
for which  our $\chi_{\vec R}$ are the common eigenfunctions.

\subsection{Generalized Cauchy formula}

A central role in applications of character theory is played by a rather elementary Cauchy formula for the bilinear sum of characters, see \cite{Cauchy} for a brief survey. It is therefore important that it has a direct counterpart for the multilinear sum related to the tensorial characters:
\be\label{ttau}
\boxed{
\begin{array}{c}
\displaystyle{
\sum_{R_1,\ldots,R_r}C_{_{R_1\ldots R_r}} \prod_{m=1}^r\chi_{R_m}\{p^{(m)}\}\stackrel{(\ref{CG})}{=}\sum_\Delta{1\over z_\Delta}\prod_{m=1}^r\left(\sum_{R_m}\psi_{_{R_m}}(\Delta)
\chi_{_{R_m}}\{p^{(m)}\}\right)=}\\
\\
\displaystyle{
=\sum_\Delta{1\over z_\Delta}\prod_{m=1}^rp^{(m)}_\Delta=\exp\left(\sum_k
{\prod_{m=1}^rp^{(m)}_k\over k}\right)}
\end{array}
}
\ee
where we used
\be
\sum_R\psi_R(\Delta)\chi_R\{p\}=p_\Delta
\ee
which follows from the orthogonality condition
\be
\sum_R\psi_R(\Delta)\psi_R(\Delta')=z_\Delta\delta_{R,R'}
\ee
and also used the fact that
\be
\sum_\Delta{p_\Delta\over z_\Delta}=\sum_n\chi_{[n]}\{p\}=\exp\left(\sum_k{p_k\over k}\right)
\ee
the last equality being nothing but the ordinary Cauchy formula.

Partition function is a rather straightforward deformation of the l.h.s. of (\ref{ttau}), which provides a tensorial generalization of KP/Toda tau-function, which we describe elsewhere \cite{IMMint}:
\be
Z_r\{p^{(i)}\}: = \sum_{R_1,\ldots,R_r} \prod_j\chi_{R_j}\{p^{(j)}\}\cdot\left<\chi_{_{R_1,\ldots,R_r}}  \right>
\stackrel{(\ref{Ga})}{=}\sum_{R_1,\ldots,R_r}C_{_{R_1\ldots R_r}} \prod_{m=1}^r\left(\chi_{R_m}\{p^{(m)}\}\cdot\prod_{(i,j)\in R_m} (N_m+i-j)\right)
\ee

\section{$\hat W$-operators
\label{Wopers}}

\subsection{Cut-and-join operators of \cite{MMN}}

For the usual Schur characters, one can construct a system of commuting differential operators $\hat W_{\!_\Delta}$ that have these characters as a common system of their eigenfunctions, the eigenvalue of
these operators being \cite{MMN}
\be
\hat W_{\!_\Delta} \chi_R = {\psi_{R}(\Delta)\over d_R}\cdot \chi_R
\label{evW}
\ee
The operators are called generalized cut and join operator \cite{MMN}, and one can construct them in terms of invariant matrix derivatives:
\be
\hat W_{\!_\Delta} = :\prod_i \hat D_{\delta_i}:
\label{Wops}
\ee
and
\be
\hat D_k = \Tr (M \p_{M})^k
\ee
where $M$ is a matrix.
The normal ordering in (\ref{Wops}) implies that all the derivatives $\p_M$
stand to the right of all $M$.
Since $W_\Delta$ are ``gauge"-invariant matrix operators, and we apply them only to
gauge invariants, they can be realized as differential operators in
the time-variables $P_k = \Tr M^k$ \cite{MMN}.
In particular, the simplest cut-and-join operator $\hat{W}_{[2]}$, \cite{GD} is
\be\label{W2}
\hat{ W}_{[2]} ={1\over 2} \sum_{a,b} \Big((a+b)P_aP_b\p_{a+b} + abP_{a+b}\p_a\p_b\Big)
\ee
These operators act on the characters considered as functions of times $p_k$
so that (\ref{evW}) reads as
\be
\hat W_{\!_\Delta} \chi_R\{p_k\} = {\psi_{R}(\Delta)\over d_R}\cdot \chi_R\{p_k\}
\ee
Note that this formula can be extended to the case of $|\Delta|\ne |R|$.
In this case, the formula looks like
\be
\hat W_{\!_\Delta} \chi_R(p_k) = \left\{\begin{array}{cc}
0&\hbox{for }|\Delta|>|R|\\
&\\
C^r_{|R|-|\Delta|+r}\displaystyle{1\over d_R}
\psi_{R}([\Delta,\underbrace{1\ldots 1}_{|R|-|\Delta|}])\cdot \chi_R(p_k)&\hbox{for }|\Delta|
\le |R|
\end{array}\right.
\label{Wexten}
\ee
where $C_n^k:={n!\over (n-k)!k!}$ are the binomial coefficients,
and $r$ is the number of unit length cycles in the Young diagram $\Delta$.

\subsection{$\hat W$-operators in rainbow tensor models}

As a direct generalization of (\ref{Wops}), one can associate a $\hat W$-operator
with any ${\cal K}_{\vec\sigma}$:
\be
\hat{\cal W}_{\vec\sigma}:=\ : {\cal K}_{\vec\sigma}\Big(\bar M
\longrightarrow {\partial\over\partial M }\Big): \
\ee
where normal ordering means that all $M$-derivatives stand to the right of all $M$.
Then, in the abbreviated notation of (\ref{arr}),
\be
\hat{\cal W}_{\vec\sigma} {\cal K}_{\vec\sigma'}
= \prod_{p=1}^n M_{\vec a_p} \frac{\p}{\p M_{\vec a_{\vec\sigma(p)}}}\
\prod_{p=1}^n M_{\vec a'_p}\bar M_{\vec a'_{\vec\sigma'(p)}}
= \sum_{\gamma \in S_n} {\cal K}_{\vec\sigma\circ\vec\sigma'\circ\gamma}
\label{calWK}
\ee
because $M$ derivatives are non-zero if the set of indices $\{\vec a_{\vec\sigma(p)}\}$
coincides with the set $\{a'_p\}$, i.e. the indices themselves are equal modulo
some $k$-independent permutation $\gamma\in S_n$, i.e.
$a^{(k)}_{\sigma_k(p)}=a'^{(k)}_{\gamma_(p)}$.
In the simplest example of $r=1$ and $n=2$,
{\footnotesize
\be
{\cal W}_\sigma{\cal K}_{\sigma'} =
\sum_{a,b=1}^N M_aM_b \frac{\p^2}{\p M_{\sigma(a)}\p M_{\sigma_b}}\
\sum_{c,d=1}^N M_cM_d \,\bar M^{\sigma'(c)}\bar M^{\sigma'(d)}
= \sum_{a,b,c,d=1}^N
\Big(\delta_{c}^{\sigma(a)}\delta_{d}^{\sigma(b)}
+ \delta_{c}^{\sigma(b)}\delta_{d}^{\sigma(a)}\Big)
\cdot M_aM_b\bar M^{\sigma'(c)}\bar M^{\sigma'(d)}
= \nn \\
= \sum_{a,b=1}^N M_aM_b \Big(\bar M^{\sigma'\circ\sigma(a)}\bar M^{\sigma'\circ\sigma(b)}
+ \bar M^{\sigma'\circ\sigma(b)}\bar M^{\sigma'\circ\sigma(a)}\Big)
= \sum_{\gamma \in S_2}
M_aM_b\, \bar M^{\sigma'\circ\sigma\circ\gamma(a)}\bar M^{\sigma'\circ\sigma\circ\gamma(b)}
=\sum_{\gamma \in S_2} {\cal K}_{\sigma'\circ\sigma\circ\gamma} \ \ \ \ \ \ \ \ \
\nn
\ee
}

In fact, ${\cal K}$ and thus $\hat{\cal W}$ are invariant w.r.t. the common multiplication
of all $n$ permutations $\sigma_k$ by a common $\sigma$, what allows to eliminate one
of them,
\be
{\cal K}_{\vec\sigma}
=  {\cal K}_{\sigma_1,\ldots,\sigma_r}
= {\cal K}_{\sigma\circ\sigma_1,\ldots,\sigma\circ\sigma_r}
= {\cal K}_{id,\sigma_1^{-1}\circ\sigma_2,\ldots,\sigma_1^{-1}\circ\sigma_r}
\ee
moreover, there is still a freedom in common {\it conjugation} of the remaining
$r-1$ permutations:
\be
{\cal K}_{id,\sigma_2,\ldots,\sigma_r} = {\cal K}_{id,\gamma\circ\sigma_2\gamma^{-1},\ldots,
\gamma\circ\sigma_r\gamma^{-1}}
\ee
In the case of RCM ($r=2$), this means that ${\cal K}$ depend only on the conjugation class
of $\sigma_1^{-1}\circ\sigma_2$, i.e. on a Young diagram, but for $r>2$ the
classification is more complicated, see \cite{IMMarist} for details.

For our present purposes, we just need to eliminate $\sigma_1$ and $\sigma_1'$ from
(\ref{calWK}).
For example, for $r=2$
\be
\hat {\cal W}_{id,\sigma}{\cal K}_{id, \sigma'}
= \sum_{\vec a,\vec b,\vec c,\vec d} M_{a_1b_1}\ldots M_{a_nb_n}
\frac{\p^n}{\p M_{a_1b_{\sigma(1)}}\ldots \p M_{a_nb_{\sigma(n)}}}
\ M_{c_1d_1}\ldots M_{c_nd_n} \bar M^{c_1d_{\sigma'(1)}}\ldots\bar M^{c_nd_{\sigma'(n)}}
= \nn \\
= \sum_{\vec a,\vec b,\vec c,\vec d} M_{a_1b_1}\ldots M_{a_nb_n}
\,\bar M^{c_1d_{\sigma'(1)}}\ldots\bar M^{c_nd_{\sigma'(n)}}
\sum_{\gamma\in S_n} \prod_{p=1}^n
\delta_{c_p}^{a_{\gamma(p)}}\delta_{d_p}^{b_{\sigma\circ\gamma(p)}}
= \sum_{\gamma\in S_n} {\cal K}_{\gamma,\sigma'\circ\sigma\circ\gamma}
= \nn \\
= \sum_{\gamma\in S_n} {\cal K}_{id,\gamma^{-1}\circ \sigma'\circ\sigma\circ\gamma}
= n!\cdot {\cal K}_{id,\sigma'\circ\sigma}
\label{WK2}
\ee
This remains true for arbitrary $r$:
\be
(\ref{calWK})\ \Longrightarrow \
\hat{\cal W}_{id,\vec\sigma}{\cal K}_{id,\vec\sigma'}
=   n!\cdot {\cal K}_{id, \vec\sigma\circ\vec\sigma'}
\label{WKid}
\ee
where now $\vec\sigma$ denotes a set of $r-1$ permutations from $S_n$.

We can now apply the $\hat {\cal W}$-operator to our $\chi$:
\be
\hat {\cal W}_{ \vec\sigma}\,\chi_{_{\vec R}} =
\frac{1}{n!}\sum_{\vec\sigma\in S_n}
\Psi_{\vec R}(\vec\sigma')\hat W_{ \vec\sigma} {\cal K}_{\vec\sigma'}
= \frac{1}{n!}
\sum_{\gamma,\vec\sigma' \in S_n} \Psi_{\vec R}(\vec\sigma')
{\cal K}_{\vec\sigma\circ\vec\sigma'\circ\gamma}
= \frac{1}{n!}
\sum_{\gamma,\vec\sigma' \in S_n} \Psi_{\vec R}(\vec\sigma^{-1}\circ\vec\sigma\circ\gamma^{-1})
{\cal K}_{\vec\sigma'}
\ee
where $\Psi_{\vec R}(\vec\sigma) := \prod_{p=1}^n \psi_{R_i}(\sigma_i)$.
This expression can be simplified a little, at expense of breaking explicit $S_n$ symmetry,
by using (\ref{WKid}) instead of (\ref{calWK}) and (\ref{chidef3}) instead of (\ref{charArist}):
\be
\hat{\cal W}_{id,\sigma_2,\ldots,\sigma_r}\, \chi_{_{R_1,\ldots,R_r}} =
  \frac{1}{n!}\sum_{\sigma'_1,\ldots,\sigma'_r\in S_n}
\psi_{R_1}(\sigma'_1)\ldots\psi_{R_r}(\sigma'_r)
\hat{\cal W}_{id,\sigma_2,\ldots,\sigma_r}\,{\cal K}_{\sigma'_1,\sigma'_2,\ldots,\sigma'_r}
= \nn \\
= \frac{1}{n!}\sum_{\sigma'_1,\ldots,\sigma'_r\in S_n}
\psi_{R_1}(\sigma'_1)\psi_{R_2}(\sigma'_1\circ\sigma'_2)\ldots\psi_{R_r}(\sigma'_1\circ\sigma'_r)
\hat{\cal W}_{id,\sigma_2,\ldots,\sigma_r}\,{\cal K}_{id,\sigma'_2,\ldots,\sigma'_r}
= \nn \\
\ \stackrel{(\ref{WKid})}{=} \
\sum_{\sigma'_1,\ldots,\sigma'_r\in S_n} \psi_{R_1}(\sigma'_1)
\psi_{R_2}(\sigma'_1\circ\sigma'_2)\ldots\psi_{R_r}(\sigma'_1\circ\sigma'_r)\,
{\cal K}_{id,\sigma_2\circ\sigma'_2,\ldots,\sigma_r\circ\sigma'_r}
\ee

\subsection{RCM, $r=2$
\label{RCMW}}

In this case, we get just
\be\label{Wed}
\hat{\cal W}_{id,\sigma} \chi_{_{R_1,R_2}}
= \sum_{\sigma'_1, \sigma'_2\in S_n} \psi_{R_1}(\sigma'_1)
\psi_{R_2}(\sigma'_1\circ\sigma'_2)\,
{\cal K}_{id,\sigma \circ\sigma'_2}
\ \stackrel{(\ref{orth})}{=} \
\frac{\delta_{R_1,R_2}}{d_{R_1}} \sum_{\sigma_2'} \psi_{R_1}(\sigma_2')\,
{\cal K}_{id,\sigma \circ\sigma'_2}
= \nn\\
=\frac{\delta_{R_1,R_2}}{d_{R_1}} \sum_{\sigma_2'} \psi_{R_1}(\sigma ^{-1}\circ\sigma_2')\,
{\cal K}_{id,\sigma'_2} =
\frac{\delta_{R_1,R_2}}{d_{R_1}}   \cdot \frac{\psi_{R_1}(\sigma)}{|R_1|!\cdot d_{R_1}}
\sum_{\sigma'_2} \psi_{R_1}(\sigma_2')\,{\cal K}_{id,\sigma_2'}
\stackrel{(\ref{chivsodchi})}{=} \frac{\psi_{R_1}(\sigma)}{d_{R_1}}\cdot \chi_{_{R_1,R_2}}
\ee
To check the next to the last transition,
one can substitute ${\cal K}_{id,\gamma}$
by arbitrary $\psi_Q(\gamma)$ (make a Fourier transform)
and then use the orthogonality relations:
\be
\sum_{\gamma\in S_n} \psi_R(\sigma^{-1}\circ\gamma) \psi_Q(\gamma)
\ \stackrel{(\ref{orth})}{=}\
\frac{\psi_R(\sigma^{-1})}{d_R}\,\delta_{QR} =
\frac{\psi_R(\sigma)}{d_R}\,\delta_{QR}
\ \stackrel{(\ref{orth1})}{=}\
 \frac{\psi_R(\sigma)}{d_R\cdot |R|!}\,\sum_\gamma \psi_R(\gamma)\psi_Q(\gamma)
\ee
For the validity of this trick, it is important that both ${\cal K}_{id,\gamma}$
and $\psi_Q(\gamma)$ depend only on the conjugation class of $\gamma$,
so that above transform is actually invertible.
We used also the fact that $\sigma^{-1}$ and $\sigma$ belong to the same conjugation class,
so that $\psi_R(\sigma^{-1})=\psi_R(\sigma)$.

Thus, we obtain that
\be\boxed{
\hat {\cal W}_{\sigma}\chi_{R_1R_2}=\lambda_{R_1R_2}^{\sigma}\cdot\chi_{R_1R_2}
}\label{Wef}
\ee
where the eigenvalues are
\be
\lambda_{R_1R_2}^{\sigma}={\psi_{R_1}(\sigma)\over d_{R_1}}
\ee

\subsection{Aristotelian model, $r=3$}

In the case of $r=3$ we have:
\be
\hat {\cal W}_{\sigma_1\sigma_2}\,\chi_{R_1R_2R_3}
\ \stackrel{(\ref{charArist})}{=}\
\frac{1}{n!}\sum_{\{\gamma_1,\gamma_2,\gamma_3\}\in S_n}
\psi_{R_1}(\gamma_1)\psi_{R_2}( \gamma_2) \psi_{R_3}(\gamma_3)\
\hat W_{\sigma_1\sigma_2}\,{\cal K}_{\gamma_1, \gamma_2,\gamma_3}
= \nn \\
=\frac{1}{n!}\sum_{\{\gamma_1,\gamma_2,\gamma_3\}\in S_n}
\psi_{R_1}(\gamma_1)\psi_{R_2}( \gamma_2) \psi_{R_3}(\gamma_3)\
\hat W_{\sigma_1\sigma_2}\,{\cal K}_{id, \gamma_1^{-1}\circ\gamma_2,\gamma_1^{-1}\circ\gamma_3}
= \nn \\
=\frac{1}{n!}\sum_{\{\gamma_1,\gamma_2,\gamma_3\}\in S_n}
\psi_{R_1}(\gamma_1)\psi_{R_2}(\gamma_1\circ\gamma_2) \psi_{R_3}(\gamma_1\circ\gamma_3)
\hat W_{\sigma_1\sigma_2}{\cal K}_{id, \gamma_2,\gamma_3}
\label{Wdef3}
\ee
If sizes $|\sigma_i|$ are equal to $|R_i|$,
then, as direct generalization of (\ref{WK2}),
the $\hat W$-operator acts as averaging over the permutation group $S_n$:
\be
\hat {\cal W}_{\sigma_1\sigma_2}{\cal K}_{id, \gamma_2,\gamma_3}
= n!\cdot {\cal K}_{id, \gamma_2\circ\sigma_1,\gamma_3\circ\sigma_2}
=\sum_{\gamma\in S_n}{\cal K}_{id,\gamma\circ\gamma_2
\circ\sigma_1\circ\gamma^{-1},\gamma\circ\gamma_3 \circ\sigma_2\circ\gamma^{-1}}
\label{Wdef3a}
\ee
In the last transition, we used invariance of operators ${\cal K}_{id, \gamma_2,\gamma_3}$
under conjugation,
see footnote \ref{var}.

\subsection{$\hat W$-operators in $\xi$-variables}

At the last step in (\ref{WK2}), we used the
fact that the operators ${\cal K}_{id, \gamma}$ are invariant w.r.t.  the conjugation
by arbitrary $\gamma'$.
However, we can ignore this symmetry and again consider the variables
$\xi_{\sigma}$ without this additional invariance.
In particular, the $\hat W$-operators
can be realized as differential operators in these variables,
\be
\hat W_{\sigma}=\sum_{\gamma,\gamma'\in S_n}\xi_{\gamma'\circ\gamma \circ\sigma
\circ\gamma'^{-1}}{\partial\over\partial\xi_{\gamma}}
\ee
Such generalized cut and join operators are linear in contrast with those in $p$-variables.
As usual, formulas in $\xi$-variables are looking simpler, for the price of
enlarging the space of variables.

\subsubsection{The case of RCM, $r=2$}
We can re-deduce (\ref{Wed}) in $\xi$-variables: from (\ref{chivsxi})
\be
\!\!
\hat W_{\sigma}\tchi_R  =
\sum_{\gamma,\gamma'\in S_n}\xi_{\gamma'\circ\gamma \circ\sigma
\circ\gamma'^{-1}}{\partial\over\partial\xi_{\gamma}}
\sum_{\gamma\in S_n}\psi_{R}(\gamma)\cdot{\xi}_{\gamma}
= \sum_{\gamma,\gamma'\in S_n} \psi_{R}(\gamma)\cdot\xi_{\gamma'\circ\gamma \circ\sigma
\circ\gamma'^{-1}}
%= \nn \\
= \sum_{\gamma,\gamma'\in S_n} \psi_{R}(\gamma\circ\sigma^{-1})
\cdot\xi_{\gamma'\circ\gamma \circ\gamma'^{-1}}
= \nn \\
\!\!\!\!\!\!\!\!\!\!\!\!
= \frac{\psi_R(\sigma)}{d_R\cdot n!}
\sum_{\gamma\gamma'\in S_n}\psi_{R}(\gamma)\cdot\xi_{\gamma'\circ\gamma \circ\gamma'^{-1}}
= \frac{\psi_R(\sigma)}{d_R\cdot n!}
\sum_{\gamma\gamma'\in S_n}\psi_{R}(\gamma'^{-1}\circ\gamma\circ\gamma')\cdot\xi_{\gamma }
= \frac{\psi_R(\sigma)}{d_R}
\sum_{\gamma\in S_n}\psi_{R}(\gamma)\cdot{\xi}_{\gamma}
=\frac{\psi_R(\sigma)}{d_R}\cdot \tchi_R
\ \
\nn
\ee
The transition between the two lines can  be explained
just by the same trick: despite, in variance with ${\cal K}_{id,\gamma}$,
the variables $\xi_\gamma$
are not supposed to be invariant under conjugations,
thus they could not be just substituted
by invariant $\psi_Q(\gamma)$,
the sums $\sum_{\gamma'\in S_n} \xi_{\gamma'\circ\gamma \circ\sigma\circ\gamma'^{-1}}$
{\it are} invariant and can be substituted so that the trick can be used.

If now one chooses $\sigma$ that labels $\hat W$-operators, and $R_1,R_2$ that label
the character belonging to different symmetric groups, $S_n$ and $S_m$ correspondingly,
the property (\ref{Wef}) still persists. It is clear that, when $n>m$,
$\lambda_{R_1R_2}^{\sigma}=0$. Otherwise, one has to extend permutations from $S_n$
to those from $S_m$ adding trivial cycles so that
\be
\lambda_{R_1R_2}^{\sigma}=\left\{\begin{array}{cc}
0&\hbox{for }n>m\\
&\\
C^{p}_{m-n+p}\ {\psi_{R_1}\big(\sigma\,(\ )^{m-n}\big)\over d_{R_1}}&\hbox{for }n\le m
\end{array}\right.
\label{Wexten1}
\ee
where $(\ )^k$ means $k$ trivial cycles added to the permutation,
and $p$ is the number of trivial cycles in the permutation $\sigma$.
This is a counterpart of the original extension formula (\ref{Wexten}).

\subsubsection{Aristotelian model, $r=3$}

In this case,
\be
\hat W_{\sigma_1\sigma_2}=\sum_{\gamma,\gamma_1\gamma_2\in S_n}
\xi_{\gamma\circ\gamma_2 \circ\sigma_1\circ\gamma^{-1},\gamma\circ\gamma_3
\circ\sigma_2\circ\gamma^{-1}}{\partial\over\partial\xi_{\gamma_1\gamma_2}}
\label{xixi3}
\ee
One can check for symmetric groups $S_n$ with small $n$ that,
when all sizes $|\sigma_i|=|R_i|=n$,
\be\boxed{
\hat W_{\sigma_1\sigma_2}\tchi_{R_1R_2R_3}
=\lambda_{R_1R_2R_3}^{\sigma_1\sigma_2}\cdot\tchi_{R_1R_2R_3}
}\label{Wef3}
\ee
where the eigenvalues are
\be
\boxed{
\bar\lambda_{R_1R_2R_3}^{\sigma_1\sigma_2}
\frac{
\sum_{\gamma\in S_n}\psi_{R_1}(\gamma)\psi_{R_2}(\gamma\circ\sigma_1)
\psi_{R_3}(\gamma\circ\sigma_2)}{C_{R_1R_2R_3}}=
\frac{C^{\sigma_1,\sigma_2}_{R_1R_2R_3}}{C_{R_1R_2R_3}}\ \ \ \hbox{at }C_{R_1R_2R_3}\ne 0
}
\label{lambda123}
\ee
since otherwise, when $C_{R_1R_2R_3}=0$, $\tchi_{R_1R_2R_3}=0$.
Now one could try to repeat the trick that was used in the proof of (\ref{orth2}) and try to prove (\ref{Wef3}), i.e. that
\be\label{C2}
C_{R_1R_2R_3}\sum_{\gamma,\gamma'}\psi_{R_1}(\gamma')\psi_{R_2}(\gamma'\circ\gamma_1\circ\gamma\circ\sigma_1\circ\gamma^{-1})
\psi_{R_3}(\gamma'\circ\gamma_2\circ\gamma\circ\sigma_2\circ\gamma^{-1})=
C^{\sigma_1\sigma_2}_{R_1R_2R_3}C^{\gamma_1\gamma_2}_{R_1R_2R_3}
\ee
Indeed, one can make a ``Fourier" transform of this formula with the kernel $C^{\sigma_1\sigma_2}_{Q_1Q_2Q_3}$ which gives for its l.h.s.
\be
C_{R_1R_2R_3}\sum_{\sigma_1\sigma_2}C^{\sigma_1\sigma_2}_{Q_1Q_2Q_3} \sum_{\gamma,\gamma'}\psi_{R_1}(\gamma')\psi_{R_2}(\gamma'\circ\gamma_1\circ\gamma\circ\sigma_1\circ\gamma^{-1})
\psi_{R_3}(\gamma'\circ\gamma_2\circ\gamma\circ\sigma_2\circ\gamma^{-1})=\nonumber\\
=C_{R_1R_2R_3}{C^{\gamma_1\gamma_2}_{Q_1Q_2Q_3}
\over d_{R_1}d_{R_2}d_{R_3}}\delta_{R_1Q_1}\delta_{R_2Q_2}\delta_{R_3Q_3}
\ee
and the same for the r.h.s.:
\be
\sum_{\sigma_1\sigma_2}C^{\sigma_1\sigma_2}_{Q_1Q_2Q_3}C^{\sigma_1\sigma_2}_{R_1R_2R_3}C^{\gamma_1\gamma_2}_{R_1R_2R_3}
={C^{\gamma_1\gamma_2}_{Q_1Q_2Q_3}C_{R_1R_2R_3}
\over d_{R_1}d_{R_2}d_{R_3}}\delta_{R_1Q_1}\delta_{R_2Q_2}\delta_{R_3Q_3}
\ee
Thus, these two formulas coincide. Unfortunately, this does not prove (\ref{C2}), since, though the l.h.s. of (\ref{C2}) does not change upon simultaneous conjugation of $\sigma_1$ and $\sigma_2$ (and, similarly, $\gamma_1$ and $\gamma_2$), the Fourier transform has no non-trivial kernel only for symmetric groups $S_n$ with small enough $n<5$.\footnote{Existence of a non-trivial kernel for large enough $n$ is clear already from the fact that dimension of the space of pairs $(\sigma_1,\sigma_2)$ invariant w.r.t. the common conjugation \cite{IMMarist}, $D=\sum_{\Delta\vdash n}z_\Delta$ grows factorially with $n$, while the number of ordered triples of Young diagrams, much slower, see \cite[s.6.1.2]{IMMarist}.}
Hence, (\ref{C2}) is proved only for these groups. We, however, have checked with the computer that this formula is correct for various concrete cases in $S_5$ and $S_6$.

In this $r=3$ case, one again can choose $\sigma_i$ labelling
$\hat W$, and $R_i$ labelling the generalized character belonging to
different symmetric groups, $S_n$ and $S_m$ correspondingly.
In this case, the property (\ref{Wef3}) still persists.
It is clear that, when $n>m$,
$\lambda_{R_1R_2R_3}^{\sigma_1\sigma_2}=0$.
Otherwise, one has to extend permutations from $S_n$ to those from $S_m$
adding trivial cycles, so that
\be
\lambda_{R_1R_2R_3}^{\sigma_1\sigma_2}=\left\{\begin{array}{cc}
0&\hbox{for }n>m\\
&\\
C^{p}_{m-n+p}\bar\lambda^{\sigma_1\,(\ )^{m-n},\sigma_2\,(\ )^{m-n}}_{R_1R_2R_3}&\hbox{for }n\le m
\end{array}\right.
\ee
where $p$ is the number of points left intact under the action
of  both permutations $\sigma_1$ and $\sigma_2$.
This is an Aristotelian counterpart of (\ref{Wexten}) and (\ref{Wexten1}).

\subsubsection{Generic $r$}

Similarly, in general, when $|\sigma_i|=|R_i|=n$, one can also make a statement, which is proved only for small symmetric groups, that
\be\boxed{
\hat W_{\vec\sigma}\tchi_{\vec R}=\lambda_{\vec R}^{\vec\sigma}\cdot\tchi_{\vec R}
}\label{Wefr}
\ee
where $\vec R$ is a set of representations $R_1,\ldots,R_r$
and the eigenvalues are
\be
\bar\lambda_{\vec R}^{\vec\sigma}
=\frac{\sum_{\gamma\in S_n}\psi_{R_1}(\gamma)
\prod_{j=1}^{r-1}\psi_{R_{j+1}}(\gamma\circ\sigma_j)}
{C_{R_1,\ldots,R_r}}
= n!\cdot \frac{\sum_{\gamma\in S_n}\psi_{R_1}(\gamma)
\prod_{j=1}^{r-1}\psi_{R_{j+1}}(\gamma\circ\sigma_j)}
{\sum_{\gamma\in S_n}
\prod_{j=1}^{r}\psi_{R_{j}}(\gamma )}\ \ \ \ \hbox{at }C_{R_1,\ldots,R_r}\ne 0
\ee
In the generic case of $|\Delta|=n$ not equal to $|R|=m$,
as the generalization of (\ref{Wexten}) and (\ref{Wexten1}),
\be
\lambda_{\vec R}^{\vec \sigma}=\left\{\begin{array}{cc}
0&\hbox{for }n>m\\
&\\
C^{p}_{m-n+p}\bar\lambda^{\vec\sigma\,(\ )^{m-n}}_{\vec R}&\hbox{for }n\le m
\end{array}\right.
\ee
where $p$ is the number of points left intact under the action
of all the permutations $\sigma_i$.

\section{Conclusion}

In this paper,
we focused on the
property (\ref{charave}),
which is extremely well suited to gaining our
knowledge in the tensor case, and thus provides a solid base for bringing new progress.
Namely, the averages in the Aristotelian model of \cite{IMMarist}
are some polynomials in $N$'s,
with the properties
\begin{itemize}
\item[(i)] they are not generic;
\item[(ii)] they are tri-linear in dimensions;
\item[(iii)] at a given level, only some of the
tri-linear products appear;
\item[(iv)] these tri-linear combinations are averages of linear combinations
of ${\cal K}$-operators with appropriate symmetries;
\item[(v)] combinations of ${\cal K}$'s with the symmetries which {\it do not}
appear in the list of tri-linear combinations identically vanish: this is the reason why they do not
appear among averages;
\item[(vi)] all this remains true for arbitrary $r$, only tri-linear combinations become
$r$-linear.
\end{itemize}

All this is a direct generalization of properties of the rank $r=2$
case (rectangular complex matrix model, RCM),
where the averages are bilinear in dimensions,
allowed are only the {\it diagonal} bilinear products $D_R(N_1)D_R(N_2)$,
and an attempt to write down an operator
$\Big<{\cal K}_{R_1R_2}\Big> \sim D_{R_1}D_{R_2}$ with $R_1\neq R_2$
fails: the operator with such a symmetry vanishes.
An additional feature of RCM is that diagonal operators ${\cal K}_{R,R}=\chi_R\{P\}$,
i.e. are just characters.
This suggests that the {\bf tensorial operators
$\chi_{R_1,\ldots,R_r}(M,\bar M)$
which we construct in this way are direct counterparts of characters}.

Among the properties that they inherit are:
\begin{itemize}
\item[(a)] orthogonality;
\item[(b)] they are eigenfunctions of generalized cut-and-join operators
$\hat W$, the tensorial counterparts of those from \cite{MMN};
\item[(c)] they form a redundant basis of the operators with non-vanishing Gaussian averages.
\end{itemize}

The {\bf main mystery} is the separation of gauge-invariant tensor model operators
into {\bf two sectors}.
One sector consists of operators which resemble characters
and possess non-vanishing Gaussian averages.
It is very similar to the conventional matrix models
like RCM, and should be rather straightforward to investigate.
The second sector is its much bigger complement,
which lies in the {\it kernel} of Gaussian averages.
The quotient structure $V/W$, in the space $V$ of gauge invariant operators, where $W\neq V$ is the vector space spanned by linear combinations of operators with vanishing Gaussian averages, is a generic feature of rainbow tensor models. It is non-trivial, because $V/W$ is {\it not a linear} subspace in the linear space $V$ and can be described in different bases. The short exact sequence 
\be
0 \longrightarrow W \longrightarrow  V \longrightarrow  V/W \longrightarrow  0
\ee 
implies interesting cohomological interpretations and calls for further investigation.

A possible clue to understanding this complement of $W$
is in a remarkable relation \cite{FDTM} which
{\bf associates gauge invariant operators with Feynman diagrams
in the theory of one-rank-less}.
One of the immediate questions to address is what characterizes
the subset of Feynman diagrams associated with  the character sector.
It would be also very useful to describe in these terms
the Virasoro-like identities
and, more generally, the CJ structure introduced in \cite{IMMarist}:
the two sectors should be somehow separated, and
the recursion between different ranks $r$ should be lifted to the
level of CJ structure. In the case of Aristotelian (rank $r=3$) model, there is a third description:
in terms of the Grothendieck's dessins, which turns especially helpful in
classifying the non-character sector in the operator space.

\bigskip

To conclude, in this paper we report the
discovery of tensorial lifting of characters
and their apparent compatibility with Gaussian averaging,
which opens absolutely new perspectives to
the theory of tensor models.
There is a whole new world to explore, and it is now clear that
it can be structured at least as well as its celebrated
matrix model predecessor is. In particular, existence of this theory of tensorial characters reflects the fact that rainbow tensor models are superintegrable and exactly solvable like their well known complex matrix model ``parent" \cite{IMMint}.

\section*{Acknowledgements}

We are grateful to the referee of our paper for valuable comments and an advice.
A. Mironov is grateful for the hospitality of NITEP, Osaka City University as well as that of the Workshop New Trends in Integrable Systems 2019 held there during the period of September, 9-20.
Our work is partly supported by JSPS KAKENHI grant Number 19K03828 and OCAMI MEXT Joint Usage/Research Center on Mathematics and Theoretical Physics (H.I.), by the grant of the Foundation for the Advancement of Theoretical Physics ``BASIS" (A.Mir., A.Mor.), by  RFBR grants 19-01-00680 (A.Mir.) and 19-02-00815 (A.Mor.), by joint grants 19-51-53014-GFEN-a (A.Mir., A.Mor.), 19-51-50008-YaF-a (A.Mir.), 18-51-05015-Arm-a (A.Mir., A.Mor.), 18-51-45010-IND-a (A.Mir., A.Mor.).
The work was also partly funded by RFBR and NSFB according to the research project 19-51-18006 (A.Mir., A.Mor.).

\end{document}